\documentclass[fleqn,usenatbib]{mnras}

\usepackage{newtxtext,newtxmath}

\usepackage[T1]{fontenc}

\DeclareRobustCommand{\VAN}[3]{#2}
\let\VANthebibliography\thebibliography
\def\thebibliography{\DeclareRobustCommand{\VAN}[3]{##3}\VANthebibliography}

\usepackage{graphicx}	
\usepackage{amsmath}	
\usepackage{ragged2e}
\usepackage{rotating}

\title[Large-scale structure around MACS0600]{A complex node of the cosmic web associated with the massive galaxy cluster MACS~J0600.1-2008}

\author[L. J. Furtak et al.]{Lukas J. Furtak,$^{1}$\thanks{E-mail: \url{furtak@post.bgu.ac.il}}
Adi Zitrin,$^{1}$
Johan Richard,$^{2}$
Dominique Eckert,$^{3}$
Jack Sayers,$^{4}$
Harald Ebeling,$^{5}$
\newauthor Seiji Fujimoto,$^{6}$
Nicolas Laporte,$^{7}$
David Lagattuta,$^{8,9}$
Marceau Limousin,$^{7}$
Guillaume Mahler,$^{10}$
\newauthor Ashish K. Meena,$^{1}$
Felipe Andrade-Santos,$^{11,12}$
Brenda L. Frye,$^{13}$
Mathilde Jauzac,$^{8,9,14,15}$
\newauthor Anton M. Koekemoer,$^{16}$
Kotaro Kohno,$^{17,18}$
Daniel Espada,$^{19,20}$
Harry Lu,$^{4}$
Richard Massey$^{8,9}$
\newauthor and Anna Niemiec$^{21}$
\\
$^{1}$Department of Physics, Ben-Gurion University of the Negev, P.O. Box 653, Be’er-Sheva 84105, Israel\\
$^{2}$Universit\'{e} Lyon~1, ENS de Lyon, CNRS, Centre de Recherche Astrophysique de Lyon UMR\,5574, 69230, Saint-Genis Laval, France\\
$^{3}$Department of Astronomy, University of Geneva, Ch. d’Ecogia 16, CH-1290 Versoix, Switzerland\\
$^{4}$California Institute of Technology, 1200 East California Boulevard, Pasadena, California 91125, USA\\
$^{5}$Institute for Astronomy, University of Hawai'i, 640 N. Aohoku Place, Hilo, HI 96720, USA\\
$^{6}$Department of Astronomy, The University of Texas at Austin, Austin, TX 78712, USA\\
$^{7}$Aix Marseille Universit\'{e}, CNRS, CNES, LAM (Laboratoire d’Astrophysique de Marseille), UMR 7326, 13388 Marseille, France\\
$^{8}$Centre for Extragalactic Astronomy, Durham University, South Road, Durham DH1 3LE, UK\\
$^{9}$Institute for Computational Cosmology, Durham University, South Road, Durham DH1 3LE, UK\\
$^{10}$STAR Institute, Quartier Agora - All\'{e}e du six Ao\^{u}t, 19c B-4000 Li\`{e}ge, Belgium\\
$^{11}$Department of Liberal Arts and Sciences, Berklee College of Music, 7 Haviland Street, Boston, MA 02215, USA\\
$^{12}$Center for Astrophysics, Harvard \& Smithsonian, 60 Garden Street, Cambridge, MA 02138, USA\\
$^{13}$Steward Observatory, University of Arizona, 933 N Cherry Ave, Tucson, AZ, USA\\
$^{14}$Astrophysics Research Centre, University of KwaZulu-Natal, Westville Campus, Durban 4041, South Africa\\
$^{15}$School of Mathematics, Statistics \& Computer Science, University of KwaZulu-Natal, Westville Campus, Durban 4041, South Africa\\
$^{16}$Space Telescope Science Institute, 3700 San Martin Drive,
Baltimore, MD 21218, USA\\
$^{17}$ Institute of Astronomy, Graduate School of Science, The University of Tokyo, 2-21-1 Osawa, Mitaka, Tokyo 181-0015, Japan\\
$^{18}$ Research Center for the Early Universe, Graduate School of Science, The University of Tokyo, 7-3-1 Hongo, Bunkyo-ku, Tokyo 113-0033, Japan\\
$^{19}$Departamento de F\'{i}sica Te\'{o}rica y del Cosmos, Campus de Fuentenueva, Edificio Mecenas, Universidad de Granada, E-18071, Granada, Spain\\
$^{20}$Instituto Carlos I de F\'{i}sica Te\'{o}rica y Computacional, Facultad de Ciencias, E-18071, Granada, Spain\\
$^{21}$LPNHE, CNRS/IN2P3, Sorbonne Universit\'{e}, Universit\'{e} Paris Cit\'{e}, Laboratoire de Physique Nucl\'{e}aire et de Hautes \'{E}nergies, 75005 Paris, France
}

\date{Accepted 2024 August 9. Received 2024 August 5; in original form 2024 April 4}

\pubyear{2024}

\begin{document}
\label{firstpage}
\pagerange{\pageref{firstpage}--\pageref{lastpage}}
\maketitle

\begin{abstract}
MACS~J0600.1-2008 (MACS0600) is an X-ray luminous, massive galaxy cluster at $z_{\mathrm{d}}=0.43$, studied previously by the \textit{REionization LensIng Cluster Survey} (RELICS) and \textit{ALMA Lensing Cluster Survey} (ALCS) projects which revealed a complex, bimodal mass distribution and an intriguing high-redshift object behind it. Here, we report on the results of a combined analysis of the extended strong lensing (SL), X-ray, Sunyaev–Zeldovich (SZ), and galaxy luminosity-density properties of this system. Using new JWST and ground-based \emph{Gemini-N} and \emph{Keck} data, we obtain 13 new spectroscopic redshifts of multiply imaged galaxies and identify 12 new photometric multiple-image systems and candidates, including two multiply imaged $z\sim7$ objects. Taking advantage of the larger areal coverage, our analysis reveals an additional bimodal, massive SL structure which we measure spectroscopically to lie adjacent to the cluster and whose existence was implied by previous SL-modeling analyses. While based in part on photometric systems identified in ground-based imaging requiring further verification, our extended SL model suggests that the cluster may have the second-largest critical area and effective Einstein radius observed to date, $A_{\mathrm{crit}}\simeq2.16\,\mathrm{arcmin}^2$ and $\theta_{\mathrm{E}}=49.7\arcsec\pm5.0\arcsec$ for a source at $z_{\mathrm{s}}=2$, enclosing a total mass of $M(<\theta_{\mathrm{E}})=(4.7\pm0.7)\times10^{14}\,\mathrm{M}_{\odot}$. These results are also supported by the galaxy luminosity distribution, the SZ and X-ray data. Yet another, probably related massive cluster structure, discovered in X-rays $5\arcmin$ (1.7\,Mpc) further north, suggests that MACS0600 is part of an even larger filamentary structure. This discovery adds to several recent detections of massive structures around SL galaxy clusters and establishes MACS0600 as a prime target for future high-redshift surveys with JWST.
\end{abstract}

%

\section{Introduction} \label{sec:intro}
As the most massive gravitationally bound structures in the Universe, galaxy clusters form relatively late in cosmic history at the nodes of the cosmic web \citep[e.g.][]{zeldovich82,delapparent86} where streams of matter fall into dense cluster cores \citep[e.g.][]{klypin83,diaferio97,springel05,dubois14,ata22}. As a result, massive galaxy clusters are often not yet fully virialized when we observe them, exhibiting merging cores \citep[e.g.][]{carrasco10,jauzac14,pascale22a,diego23,diego24}, disturbed X-ray gas \citep[e.g.][]{borgani01,ebeling04,merten11} and intra-cluster light \citep[ICL; e.g.][]{durret16,ellien19}, radio relics \citep[e.g.][]{bonafede12,finner17,finner21}, as well as extended massive structures far outside their cores \citep[e.g.][]{kartaltepe08,jauzac12,jauzac18,medezinski16,furtak23c}. In light of this complexity, gravitational lensing represents the method of choice to constrain the distribution of otherwise invisible dark matter (DM) in and around evolving galaxy clusters since lensing is sensitive to the projected mass density distribution yet does not rely on the assumption of hydrostatic equilibrium \citep[e.g.][]{geller13}. In fact, multiple DM cores and sub-structure are known to often enhance and extend the strong-lensing (SL) regime \citep[e.g.][]{meneghetti03,meneghetti07a,meneghetti07b,torri04,redlich12,zitrin13a,zitrin13b}.

Massive and merging SL clusters that cover a relatively large area on the sky are particularly valuable as `cosmic telescopes' that use the gravitational magnification of the cluster to enable the study of faint objects in the background which would otherwise not be observable \citep[e.g.][]{maizy10,sharon12,monna14,coe15,coe19}. Indeed, thanks to the magnification provided by SL, cluster surveys conducted with the \textit{Hubble Space Telescope} (HST) were able to detect several hundreds of high-redshift ($z\sim6-10$) galaxies down to luminosities of $M_{\mathrm{UV}}\lesssim-13$ \citep[e.g.][]{livermore17,bouwens17,bouwens22b,ishigaki18,atek18} and stellar masses of $M_{\star}\gtrsim10^6\,\mathrm{M}_{\odot}$ \citep[e.g.][]{bhatawdekar19,kikuchihara20,furtak21,strait20,strait21}. More recently, cluster observations with JWST \citep[][]{gardner23,mcelwain23} discovered the first galaxy candidates at $z>10$ \citep[e.g.][]{adams23,atek23a,bradley23}, enabled the first rest-frame optical spectroscopy at $z>6$ \citep[e.g.][]{roberts-borsani23,williams23,hsiao24a} which confirmed the high ionization power of low-mass galaxies in the early Universe \citep[][]{atek24a}, and uncovered a new population of dust-obscured active galactic nuclei (AGN), the so-called ``little red dots'' \citep[LRDs;][see also \citealt{matthee24}]{furtak23d,furtak24b,labbe24,greene24}. Thanks to the extreme magnifications achieved when a background object crosses the critical line, it has become possible to detect and study extremely compact star clusters and even single stars out to high redshifts \citep[e.g.][]{kelly18,welch22a,welch22b,meena23a,furtak24a}. Finally, multiply-imaged supernovae \citep[e.g.][]{frye24} can be used to constrain cosmological parameters thanks to the delay in arrival time between the multiple images \citep[e.g.][]{pascale24}. Discovering new, powerful cluster lenses is therefore a crucial step in preparation for the next generations of high-redshift surveys.

Although not yet virialized, massive merging clusters are usually well defined over-densities in the cosmic web \citep[e.g.][]{geller14,geller16} and, at the same time, frequently connected to extended cosmic filaments, e.g. MACS~J0717.5+3745 \citep[e.g.][]{jauzac12,jauzac18,durret16,ellien19} or Abell~2744 \citep[][]{medezinski16,jauzac16}. The very existence of the most massive galaxy clusters such as e.g. \textit{El Gordo} \citep[e.g.][]{menanteau12,zitrin13b,molnar15,diego23,frye23}, and even more so their observed number densities, have been claimed to pose a challenge to Lambda cold dark matter ($\Lambda$CDM) cosmology \citep[e.g.][]{donahue98,mullis05,ebeling07,foley11,miller18,asencio21}. Observing galaxy clusters at the high-mass end of the hierarchical Universe and constraining their total mass is crucial, since the extreme-value statistics of the cluster mass function constrain large-scale structure formation in the Universe \citep[e.g.][]{frye23} in the same way that the most massive galaxies constrain galaxy formation \citep[e.g.][]{glazebrook24}.

In this work, we present the discovery of cluster-scale structures associated with the massive cluster lens MACS~J0600.1-2008 \citep[also known as RXC~J0600.1-2007; MACS0600 hereafter;][]{ebeling01}, which was targeted in the \textit{REionization LensIng Cluster Survey} \citep[RELICS;][]{coe19} with HST and the \textit{ALMA Lensing Cluster Survey} \citep[ALCS;][]{kohno23,fujimoto23} program with the \textit{Atacama Large Millimeter/sub-millimeter Array} (ALMA). The cluster was previously known for hosting a Type-Ia supernova \citep{coe19,golubchik23} and the quintuple image of an ALMA-detected galaxy at $z=6.0719\pm0.0004$ \citep{laporte21b,fujimoto21}, which we recently observed at high resolution with JWST \citep[][]{fujimoto24}. We derive an improved SL model for MACS0600 based on new observations and, for the first time, model the extended mass distribution of the cluster.

This paper is organized as follows: In Section~\ref{sec:data} we present the data sets used in this work. Section~\ref{sec:SL-modeling} describes our SL analysis methods. The results are then presented in Section~\ref{sec:result} and discussed in Section~\ref{sec:discussion}. Finally, we conclude this paper in Section~\ref{sec:conculsion}. Throughout this work, we assume a standard flat $\Lambda$CDM cosmology with $H_0=70$ km s$^{-1}$ Mpc$^{-1}$, $\Omega_{\Lambda}=0.7$, and $\Omega_\mathrm{m}=0.3$. All magnitudes are quoted in the AB system \citep{oke83}, and all quoted uncertainties represent $1\sigma$ confidence limits unless stated otherwise.

\section{Target and observations} \label{sec:data}

\begin{figure*}
    \centering
    \includegraphics[width=\textwidth]{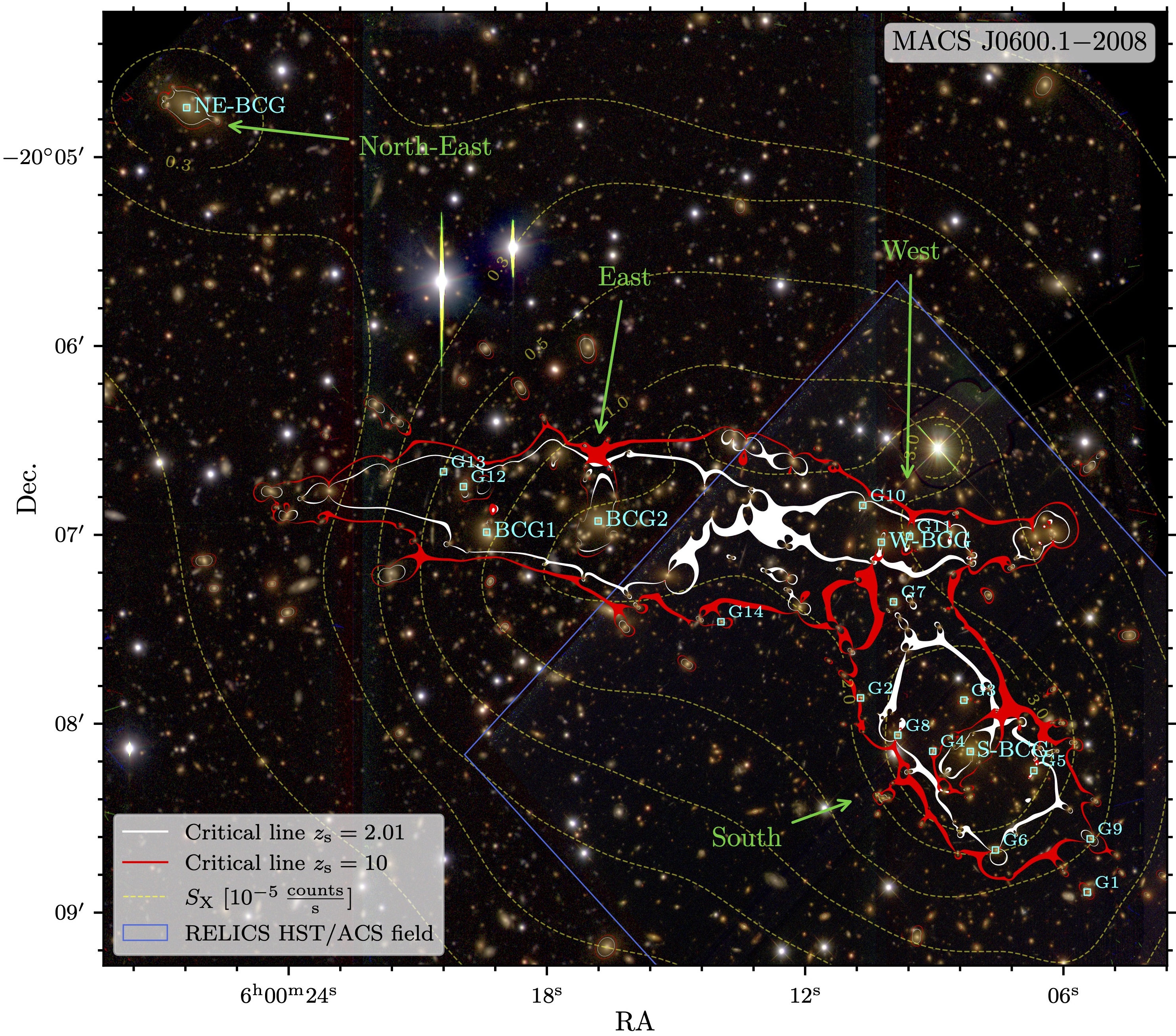}
    \caption{Combined HST/ACS and \textit{Gemini-N}/GMOS composite-color image (Blue: \textit{g}+F435W, Green: \textit{r}+F606W, Red: \textit{i}+F814W) of MACS0600. The BCGs of the cluster (section~\ref{sec:cluster_members}) and a number of cluster member galaxies of interest (see Tab.~\ref{tab:cluster_members}) are shown in cyan. Overlaid are the critical curves of our SL model (section~\ref{sec:SL_result}) for $z_{\mathrm{s}}=2.01$ (corresponding to system~1) and $z_{\mathrm{s}}=10$ in white and red, respectively. The green arrows designate the various cluster sub-structures analyzed in this work. The X-ray surface brightness (see sections~\ref{sec:ancillary} and~\ref{sec:X-ray_gas}, and Fig.~\ref{fig:X-ray_gas}) is overlaid as yellow contours in steps of $[0.1, 0.2, 0.3, 0.5, 1.0, 2.0, 3.0]\times10^{-5}\,\frac{\mathrm{counts}}{\mathrm{s}}$. The RELICS HST/ACS field-of-view (section~\ref{sec:RELICS}) is shown in blue. We nickname MACS0600 `Anglerfish cluster' analogous to the deep-sea creature of which often only one part is immediately seen and the bulk remains hidden. A full-resolution (0.06\arcsec/pix), fully vectorized version of this figure is included in the online supplementary material of this paper.}
    \label{fig:crit-curves}
\end{figure*}

\begin{figure}
    \centering
    \includegraphics[width=\columnwidth]{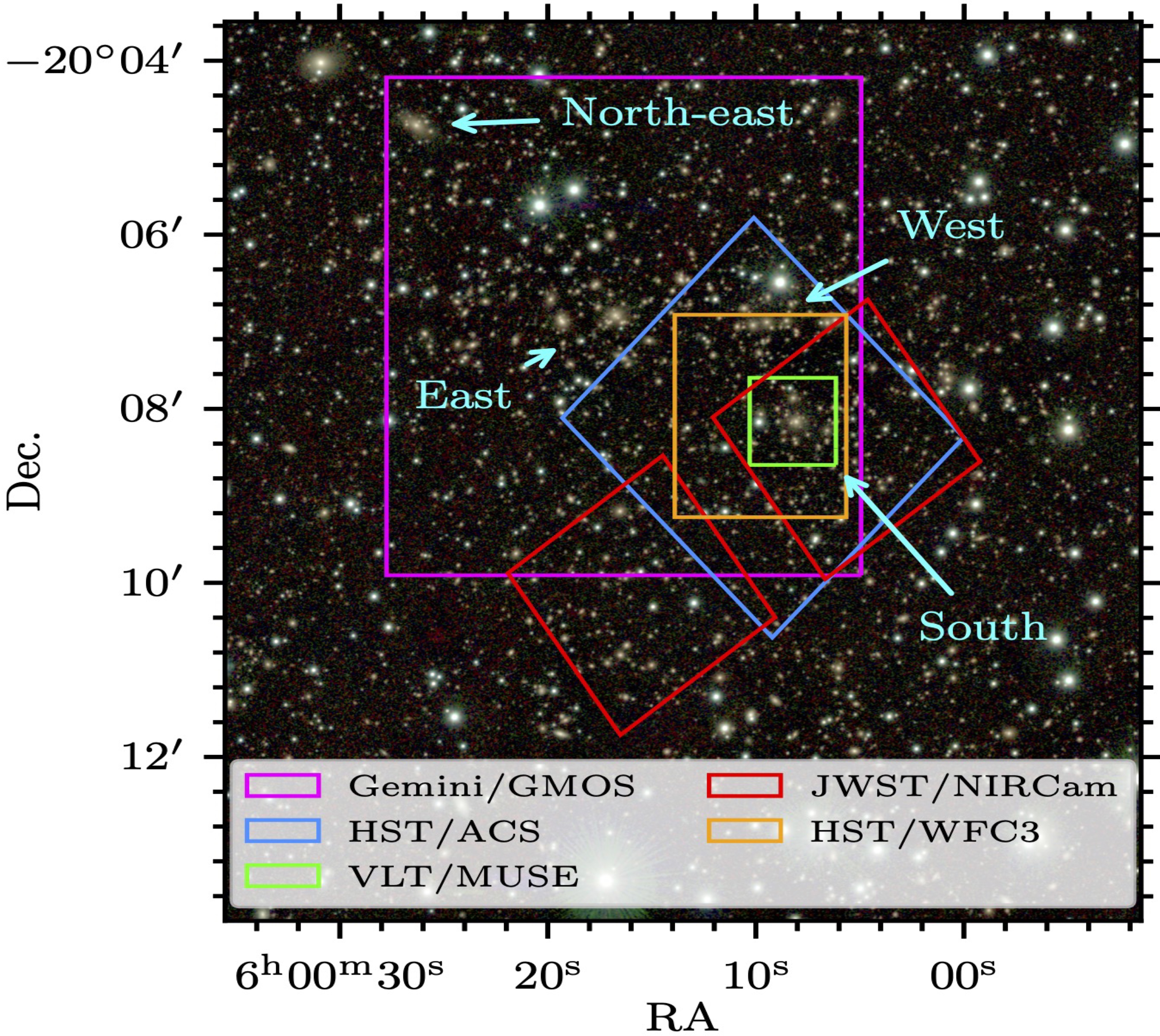}
    \caption{Cutout of a VIRCAM color-composite image (blue: \textit{Y}-band, green: \textit{J}-band, red: \textit{Ks}-band) overlaid with the field-of-views of the various observations of the MACS0600 used in this work (see sections~\ref{sec:gemini}, \ref{sec:JWST}, \ref{sec:RELICS}, \ref{sec:muse} and~\ref{sec:ancillary}).}
    \label{fig:fields}
\end{figure}

\begin{table}
    \centering
    \caption{Summary of observations used in this work. The pointings of the most important data sets can be seen in Fig.~\ref{fig:fields}.}
    \begin{tabular}{lcr}
    \hline
     Instrument             &   Exposure time       &   Reference\\\hline
     \multicolumn{3}{c}{\textit{Imaging}}\\\hline
     \textit{Gemini-N}/GMOS &   40\,min             &   This work\\
     JWST/NIRCam            &   $0.5-1.5$\,h        &   \citet{fujimoto24}\\
     HST/ACS \& WFC3        &   $0.5-1$\,orbits     &   RELICS; \citet{coe19}\\
     VISTA/VIRCAM           &   45\,min             &   GCAV\\
     UH 2.2\,m telescope    &   240\,s              &   \citet{repp18}\\\hline
     \multicolumn{3}{c}{\textit{Spectroscopy}}\\\hline
     \textit{Gemini-N}/GMOS &   2.3\,h              &   This work\\
     VLT/MUSE               &   7.3\,h              &   This work\\
     \textit{Keck}/DEIMOS   &   1.0\,h              &   This work\\\hline
     \multicolumn{3}{c}{\textit{Ancillary}}\\\hline
     \textit{XMM-Newton}    &   42\,ks              &    CHEX-MATE\\
     ACT                    &   -                   &     \citet{mallaby-kay21}\\\hline
    \end{tabular}
    \label{tab:observations}
\end{table}

This study focuses on galaxy cluster structures detected north-east ward of the X-ray luminous galaxy cluster MACS0600 \citep[][]{ebeling01}, just beyond the field-of-view of the HST observations taken in the framework of the RELICS program. Note that MACS0600 was selected for RELICS precisely because of its high Sunyaev-Zel'dovich \citep[SZ;][]{sunyaev72} mass \citep{coe19} of $M_{500,\mathrm{SZ}}=10.73_{-0.54}^{+0.51}\times10^{14}\,\mathrm{M}_{\odot}$ \citep{planck16_xxvii}.

A significant mass outside of the cluster was first suspected due to difficulties in reproducing the strongly lensed multiple images in MACS0600. We therefore obtained optical imaging with the \textit{Gemini-North} telescope (\textit{Gemini-N}) and indeed find a prominent overdensity of cluster galaxies north-east of the RELICS field-of-view (see Fig.~\ref{fig:crit-curves}). Initial estimates, based on galaxy velocity dispersions (see section~\ref{sec:extended}) suggest that this structure might even be more massive than the HST-observed part of MACS0600. We therefore dub this cluster the `Anglerfish cluster', since analogous to the deep-sea creature, we have been observing only part of it all this time, thereby missing the other, perhaps larger part. From here on, we will refer to the new structure in the north-east as the `eastern' part of the cluster, the southern part of MACS0600 (formerly its center) as the `southern' structure and the structure north of it as the `western' clump, as shown in Fig.~\ref{fig:crit-curves}.

The following sections describe the observations used in this study: Our \textit{Gemini-N} observations in section~\ref{sec:gemini}, recent JWST imaging of the southern clump in section~\ref{sec:JWST}, the RELICS data set in section~\ref{sec:RELICS}, and additional ancillary spectroscopy, ground-based imaging, X-ray and sub-millimeter observations in sections~\ref{sec:DEIMOS} and~\ref{sec:ancillary}. The relative positions and coverages of the various data sets used here are shown in Fig.~\ref{fig:fields} and an overview of the observations is given in Tab.~\ref{tab:observations}.

\subsection{\textit{Gemini-N}/GMOS observations} \label{sec:gemini}
To observe the structures suspected outside of the RELICS field-of-view of MACS0600 (the `eastern' part of the cluster), we obtained broad-band imaging data with the \textit{Gemini Multi-Object Spectrograph} \citep[GMOS;][]{hook04} on \textit{Gemini-N} in three filters: \textit{g}, \textit{r} and \textit{i} (Program ID: GN-2020B-Q-903; PI: A.~Zitrin). The observations of the region north-east of MACS0600 total 40\,minutes integration time per band. The data are first reduced and co-added with the standard GMOS pipeline in the \texttt{dragons} package \citep[\texttt{v3.0.1;}][]{dragons21}, background-subtracted with \texttt{photutils} \citep[\texttt{v1.4.0};][]{photutils22} and then registered to the astrometry of the \textit{Wide-field Infrared Survey Explorer} (WISE) point-source catalog \citep[][]{wright10} using \texttt{scamp} \citep[][]{bertin06}. This last step ensures that we are using the same astrometry as the existing RELICS catalog \citep[][see section~\ref{sec:RELICS}]{coe19}. The final images achieve a pixel scale of 0.16\arcsec/pix. We run \texttt{SExtractor} \citep{bertin96} in dual-image mode, with a stack of the three GMOS bands as the detection image, to detect sources and measure photometry. The photometry is then corrected for galactic extinction using the \citet{schlafly11} dust emission maps.

Within the same program, we also obtained 2.3\,h of \textit{Gemini-N}/GMOS slit-spectroscopy of selected objects both within the MACS0600 HST field-of-view and the region in the new eastern part. We in particular targeted the suspected BCGs of the cluster structure in the east. The data were reduced and spectra extracted using the \textit{Gemini} \texttt{IRAF} data reduction package\footnote{\url{https://www.gemini.edu/observing/phase-iii/reducing-data}}. The data reduction workflow includes bias subtraction, flat-fielding and wavelength calibration using a \textsc{CuAr} lamp illuminating our mask. Flux calibration was performed using the white dwarf Wolf~1346 as a standard star. The final spectra achieve $1\sigma$-sensitivities of $9\times10^{-19}\,\frac{\mathrm{erg}}{\mathrm{s}\,\mathrm{cm}^{-2}\,\text{\AA}}$.

\subsection{JWST imaging} \label{sec:JWST}
MACS0600 has also recently been observed with JWST in cycle~1 (Program ID: GO-1567; PI: S.~Fujimoto), both with the \textit{Near Infrared Camera} \citep[NIRCam;][]{rieke23} and the \textit{Near Infrared Spectrograph} \citep[NIRSpec;][]{jakobsen22,boeker23} integral field unit \citep[IFU;][]{boeker22} targeted at the ALMA-detected $z=6.07$ arc \citep{fujimoto21,laporte21b}. The JWST observations and results are published in \citet{fujimoto24}.

For the purpose of this work, we use the NIRCam imaging, which comprises five broad-band filters: F115W, F150W, F277W, F356W and F444W. The images have exposure times of 30\,minutes in F115W and F277W, 1.4\,h in F150W, and 40\,minutes in F356W and F444W. After initial calibration with the JWST calibration pipeline \citep[][]{bushouse23}, the data, retrieved from the \texttt{Mikulski Archive for Space Telescopes} (\texttt{MAST}), were reduced and drizzled into mosaics with the \texttt{grism redshift and line analysis software for space-based spectroscopy} pipeline \citep[\texttt{grizli};][]{grizli23}. The final short-wavelength (SW; i.e. F115W and F150W) mosaics have resolutions of 0.02\arcsec per pixel and the long-wavelength (LW; F277W, F356W and F444W) mosaics of 0.04\arcsec per pixel. The footprint of the JWST imaging compared to the other available data sets can be seen in Fig.~\ref{fig:fields}. We refer the reader to \citet{fujimoto24} for more details on the reduction and processing of the NIRCam images.

\subsection{Archival RELICS HST imaging data} \label{sec:RELICS}
In addition, we have at our disposal the public data set of the RELICS program. The RELICS HST imaging data of MACS0600 comprises mosaics in 7 broad-band filters of the \textit{Advanced Camera for Survey} (ACS) and the \textit{Wide Field Camera Three} (WFC3): F435W, F606W, F814W, F105W, F125W, F140W and F160W, which are available in the RELICS repository on the \texttt{Mikulski Archive for Space Telescopes} (\texttt{MAST}). In this work we use the ACS+WFC3 catalog, also available in \texttt{MAST}, which contains HST fluxes obtained with \texttt{SExtractor} and photometric redshifts computed with the \texttt{Bayesian Photometric Redshifts} code \citep[\texttt{BPZ};][]{benitez00,coe06}. We refer the reader to \citet{coe19} for the details of data reduction, catalog assembly and photometric redshift computation.

\subsection{VLT/MUSE spectroscopy} \label{sec:muse}
Spectroscopic observations of the southern clump of MACS0600 were performed with the ESO \textit{Very Large Telescope} (VLT) \textit{Multi Unit Spectroscopic Explorer} \citep[MUSE;][]{bacon10} (Program ID: 0100.A-0792(A); PI: A.~Edge, and Program ID: 109.22VV.001(A); PI: S.~Fujimoto), in Jan.~2018 and Sep.-Dec.~2022, respectively. The 2018 observations were taken without adaptive optics, during 3 exposures of 900\,s each and are thus relatively shallow. The 2022 observations make use of the adaptive optics system (WFM-AO-N) and comprise of 18 exposures of 1284\,s each. The MUSE pointing was centered on $\alpha=$06:00:08.8 and $\delta=$-20:08:08 in both cases (see Fig.~\ref{fig:fields}).

We follow the procedure described in \citet{richard21} for all steps related to the data reduction, spectral extraction and redshift measurements. This approach is optimised for deep MUSE observations towards crowded lensing cluster fields. The basic steps of data reduction make use of version \texttt{v2.7} of the ESO data reduction software \citep{weilbacher20} (which includes correction to the heliocentric frame), but include the self-calibration and sky subtraction steps with a specifically adjusted object mask to account for the extended light envelopes of cluster member galaxies. Individual exposures are resampled onto the same cube sampling grid and then the cubes are combined with the \texttt{MPDAF} \citep{piqueras19} cube combination recipes \footnote{\url{https://mpdaf.readthedocs.io/en/latest/cubelist.html}}. Higher systematics in sky background are removed by masking specific inter-stack regions of the MUSE slicer in individual cubes. The combined cube is then processed with the \texttt{ZAP} \citep[\texttt{v2};][]{soto16} software which performs additional subtraction of sky residuals with a principle component analysis (PCA) technique. Finally, absolute astrometry of the combined data cube is calibrated against available HST images. Image quality is measured in the final data cube to be 0.59\arcsec\ FWHM at 7000\,\AA.

Following \citet{richard21}, source detection is performed using a combination of HST-based priors (from the RELICS HST catalog; see section~\ref{sec:RELICS}) and a full search for line emitters in the MUSE cube using the \texttt{MUSELET} software\footnote{\url{https://mpdaf.readthedocs.io/en/latest/muselet.html}}. MUSE spectra are extracted using the optimized algorithm from \citet{horne86}, and include a subtraction of the local background. The spectra are then run through the \texttt{MarZ} software \citep{hinton16}  to identify possible redshift solutions through cross-correlation with spectral templates. These solutions are finally inspected and each source is assessed a final redshift and a confidence level ranging from 1 to 3, where 3 is the most reliable.

\subsection{\textit{Keck}/DEIMOS spectroscopy} \label{sec:DEIMOS}
Additional multi-object spectroscopy (MOS) observations of MACS0600 were performed on Oct.~28, 2016, with the \textit{DEep Imaging Multi-Object Spectrograph} \citep[DEIMOS;][]{faber03} on the \textit{Keck-2} 10\,m telescope on Maunakea, Hawai'i. The individual MOS targets were selected from prior wide-field imaging observations of the system conducted with the University of Hawai'i (UH) 2.2\,m telescope in the \textit{V}-, \textit{R}-, and \textit{I}-band filters, by using the resulting color information of galaxies in the target field as an indicator of likely cluster membership \citep{repp18}. The spectra were taken in three exposures of 1200\,s each. Since only a single mask was designed and observed, priority was given to maximizing the number of target galaxies of a common color without attempting to achieve statistical completeness either in magnitude or color space. Using a 1\arcsec\ slit width and a central wavelength of 6300\,\AA, the instrumental setup for the DEIMOS follow-up observations combined the $600\,\frac{\mathrm{line}}{\mathrm{mm}}$ Zerodur grating with a GG455 order-blocking filter, resulting in a 1.2\,\AA\ per pixel sampling. We measured redshifts by cross-correlation with spectral templates and applied corrections to the heliocentric frame using an adaptation of the \texttt{SpecPro} package \citep{specpro11}.

\subsection{Ancillary data} \label{sec:ancillary}
Additional ancillary data available for MACS0600 include wide-field near-infrared (NIR) imaging taken with the \textit{VISTA infrared camera} \citep[VIRCAM;][]{dalton06} on ESO's \textit{Visible and Infrared Survey Telescope for Astronomy} (VISTA) in the framework of the \textit{Galaxy Clusters At VIRCAM} survey (GCAV; Program ID: 198.A-2008(E); PI: M.~Nonino). The reduced \textit{Y}-, \textit{J}- and \textit{Ks}-band images achieve $5\sigma$ depths of $23.5$\,magnitudes and their $1.5\times1.2$ degree field-of-view is optimal for detecting possible extended structures around the cluster. Initial studies of the extended cluster structures of MACS0600 were conducted using optical imaging from the UH 2.2\,m telescope in the \textit{V}-, \textit{R}- and \textit{I}-bands to depths of 240\,s each.

We also reduced and analyzed publicly available X-ray observations of MACS0600, obtained with \emph{XMM-Newton} in the framework of the \textit{Cluster HEritage project with XMM-Newton - Mass Assembly and Thermodynamics at the Endpoint of structure formation} \citep[CHEX-MATE;][]{chexmate21} program, to study the distribution of the hot intra-cluster medium (ICM) gas and the large-scale environment of the system. MACS0600 was observed by \emph{XMM-Newton} on two occasions (observation IDs:~0650381401 and~0827050601) for a total exposure time of 42\,ks. We reduced the two observations using \texttt{XMMSAS} {\texttt{v20.0}} and the \texttt{X-COP} analysis pipeline \citep{eckert17,ghirardini19}. We used the standard data reduction chains to extract clean event lists for the three \textit{European Photon Imaging Camera} \citep[EPIC;][]{struder01,turner01} detectors (MOS1, MOS2, and PN), and constructed the light-curve of the observation to filter-out time periods affected by soft proton flares. After flare filtering, the total clean observing time is of 30.2\,ks for the two MOS detectors and 23.5\,ks for the PN detector. We extract photon maps from each observation and each detector in 5 energy bands spanning the $0.5-7$\,keV band, and stacked the maps to increase their depth. The non X-ray background was modeled using a large collection of filter-wheel-closed observations and re-scaled to each observation by matching the high-energy count rates measured in the unexposed corners of the detectors. For more details on the procedure, we refer the reader to \citet{ghirardini19}.

Last, we also use mm-wave data used to map the SZ effect of the cluster, using \textit{Atacama Cosmology Telescope} (ACT) 90\,GHz and 150\,GHz maps from ACT data release 5 \citep{mallaby-kay21}, combined with \textit{Planck} data \citep{naess20}. The angular resolution of the maps is set by the ACT PSF, which has an approximate full-width at half maximum (FWHM) of 2.1\arcmin\ and 1.4\arcmin\ at 90\,GHz and 150\,GHz \citep{choi20}. To remove primary cosmic microwave background (CMB) anisotropies from the maps, we subtract the component-separated CMB map obtained from the \texttt{COMMANDER} algorithm in \textit{Planck} data release 3 \citep{planck20_iv}. Finally, note that MACS0600 was also observed in ALMA band~6 (1.1\,mm) in the framework of the ALCS program, though we do not make use of these data in this study.

\section{Gravitational lensing mass modeling} \label{sec:SL-modeling}
We model the cluster using the new implementation of the parametric SL mass modeling code by \citet{zitrin15a}, which was presented in \citet{pascale22b} and \citet{furtak23c}, often referred to as \texttt{Zitrin-analytic} in recent works. Similarly to other parametric codes \citep[e.g.][]{jullo07,oguri10b}, it models the various components with typical parametric functions. For the smooth cluster DM components we use here a pseudo-isothermal elliptical mass distribution \citep[PIEMD; e.g.][]{kassiola93,keeton01a,monna17}:

\begin{equation} \label{eq:PIEMD}
    \Sigma(\xi)=\frac{\sigma_v^2}{2G}\left(r_{\mathrm{core}}^2+\xi^2\right)^{-\frac{1}{2}},
\end{equation}

\noindent and cluster member galaxies are modeled as dual pseudo-isothermal ellipsoids \citep[dPIEs; e.g.][]{kassiola93,keeton01a,eliasdottir07,monna17}:

\begin{equation} \label{eq:dPIE}
    \Sigma(\xi)=\frac{\sigma_v^2r_{\mathrm{cut}}^2}{2G\xi(r_{\mathrm{cut}}^2-r_{\mathrm{core}}^2)}\left[\left(1+\frac{r_{\mathrm{core}}^2}{\xi^2}\right)^{-\frac{1}{2}}-\left(1+\frac{r_{\mathrm{cut}}^2}{\xi^2}\right)^{-\frac{1}{2}}\right].
\end{equation}

\noindent In both equations, $\sigma_v$ is the velocity dispersion of the profile, $r_{\mathrm{core}}$ and $r_{\mathrm{cut}}$ are core and truncation radii respectively and $\xi$ is an elliptical coordinate defined as:

\begin{equation} \label{eq:xi}
    \xi^2=\frac{x^2}{(1-\epsilon)^2}+\frac{y^2}{(1+\epsilon)^2}.
\end{equation}

The ellipticity $\epsilon$ is defined as

\begin{equation} \label{eq:epsilon}
    \epsilon=\frac{b-a}{a+b}
\end{equation}

\noindent where $a$ and $b$ are the semi-major and -minor axes respectively. Note that, in practice, the mass profiles~\eqref{eq:PIEMD} and~\eqref{eq:dPIE} are implemented following the parametrization of \citet{keeton01a} and \citet{oguri10a}. We refer the reader to \citet{furtak23c} for more details on the method.

The cluster member galaxy selection is described in section~\ref{sec:cluster_members}, the multiple images used to constrain the SL model of MACS0600 are described in section~\ref{sec:multiple_images} and finally the SL modeling setup, in section~\ref{sec:SL_setup}.

\subsection{Cluster member galaxies} \label{sec:cluster_members}

\begin{figure}
    \centering
    \includegraphics[width=\columnwidth, keepaspectratio=true]{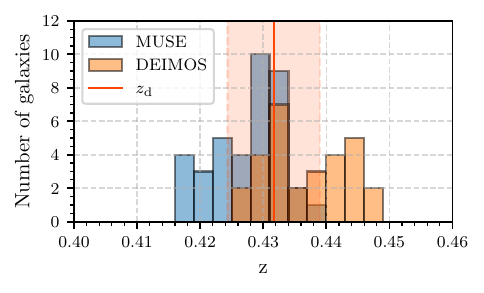}
    \caption{Spectroscopic redshift distribution of galaxies in MACS0600. The blue histogram represents the MUSE spectroscopic cluster members in MACS0600 and the orange histogram shows spectroscopic redshifts obtained with DEIMOS which extend to the eastern sub-structure. The cluster redshift $z_{\mathrm{d}}=0.432\pm0.007$ is shown in red.}
    \label{fig:cluster_members}
\end{figure}

\begin{figure}
    \centering
    \includegraphics[width=\columnwidth, keepaspectratio=true]{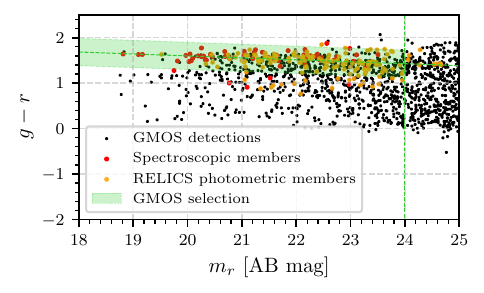}
    \caption{Color-magnitude diagram of \textit{Gemini-N}/GMOS sources (black dots). The red sequence is clearly visible in the $g-r$ color which straddles the 4000\,\AA-break at the cluster redshift $z_{\mathrm{d}}=0.432$. We show the GMOS colors of known spectroscopic cluster members in red and of HST photometry based cluster members in orange. Our GMOS red sequence selection window is shown in green}
    \label{fig:red-sequence}
\end{figure}

\begin{table*}
    \centering
    \caption{BCGs in MACS0600 cluster shown in Fig.~\ref{fig:crit-curves} and other individual galaxies whose weight was left as a free parameter in our SL model (see section~\ref{sec:SL_setup}).}
    \begin{tabular}{lcccccccc}
    \hline
    Galaxy ID                &   RA$^{\mathrm{a}}$ &   Dec.$^{\mathrm{a}}$ &   $m_r$$^{\mathrm{a}}$ &   $z_{\mathrm{spec}}$ &  $\sigma_v^{\mathrm{b}}~\left[\frac{\mathrm{km}}{\mathrm{s}}\right]$ &   $r_{\mathrm{core}}^{\mathrm{b}}~[\mathrm{kpc}]$ &   $r_{\mathrm{cut}}^{\mathrm{b}}~[\mathrm{kpc}]$ & $M_{\mathrm{tot}}~[10^{12}\,\mathrm{M}_{\odot}]$\\\hline
    \multicolumn{8}{c}{\textit{BCGs shown in Fig.~\ref{fig:crit-curves}}}\\\hline
    BCG1                     &   06:00:19.40       &   -20:06:59.18        &   $18.814\pm0.013$     &   $0.4403\pm0.0001$   &  $490_{-21}^{+16}$                                                   &   $0.375_{-0.006}^{+0.006}$                       &   $155_{-26}^{+21}$                               &  $27_{-9}^{+8}$\\
    BCG2                     &   06:00:16.81       &   -20:06:55.67        &   $19.096\pm0.013$     &   $0.4357\pm0.0002$   &  $427_{-19}^{+14}$                                                   &   $0.335_{-0.005}^{-0.005}$                       &   $131_{-22}^{+18}$                              &  $17_{-6}^{+5}$\\
    South BCG$^{\mathrm{c}}$ &   06:00:08.17       &   -20:08:08.85        &   $19.913\pm0.002$     &   $0.4266\pm0.0002$   &  $406_{-18}^{+14}$                                                   &   $0.237_{-0.005}^{+0.007~\mathrm{d}}$            &   $123_{-20}^{+18}$                              &  $15_{-5}^{+4}$\\
    West BCG $^{\mathrm{c}}$ &   06:00:10.23       &   -20:07:02.37        &   $20.018\pm0.003$     &   $0.4460\pm0.0002$   &  $341_{-85}^{+38}$                                                   &   $0.057_{-0.016}^{+0.006~\mathrm{d}}$            &   $100_{-34}^{+18}$                              &  $8_{-7}^{+4}$\\
    North-East BCG           &   06:00:26.37       &   -20:04:44.30        &   $18.831\pm0.013$     &   -                   &  $477_{-21}^{+16}$                                                   &   $0.367_{-0.005}^{+0.006}$                       &   $150_{-25}^{+20}$                              &  $25_{-8}^{+7}$\\\hline
    \multicolumn{8}{c}{\textit{Cluster galaxies with a free weight (see Fig.~\ref{fig:crit-curves})}}\\\hline
    G1                       &   06:00:05.45       &   -20:08:53.53        &   $21.499\pm0.008$     &   -                   &  $183_{-9}^{+12}$                                                    &   $0.165_{-0.008}^{+0.004}$                       &    $47_{-8}^{+7}$                                &  $1.15_{-0.40}^{+0.38}$\\
    G2                       &   06:00:10.72       &   -20:07:51.84        &   $21.028\pm0.005$     &   -                   &  $68_{-12}^{+12}$                                                    &   $0.073_{-0.011}^{+0.011}$                       &    $14_{-4}^{+4}$                                &  $0.05_{-0.03}^{+0.03}$\\
    G3                       &   06:00:08.32       &   -20:07:52.54        &   $20.417\pm0.005$     &  $0.4316\pm0.0002$    &  $184_{-30}^{+12}$                                                   &   $0.166_{-0.008}^{+0.022}$                       &    $47_{-12}^{+7}$                               &  $1.16_{-0.70}^{+0.38}$\\
    G4                       &   06:00:09.04       &   -20:08:08.78        &   $20.902\pm0.004$     &  $0.4369\pm0.0002$    &  $116_{-13}^{+9}$                                                    &   $0.113_{-0.007}^{+0.010}$                       &    $27_{-5}^{+4}$                                &  $0.26_{-0.12}^{+0.09}$\\
    G5                       &   06:00:06.69       &   -20:08:14.93        &   $20.919\pm0.005$     &  $0.4307\pm0.0002$    &  $265_{-12}^{+11}$                                                   &   $0.225_{-0.005}^{+0.004}$                       &    $73_{-12}^{+10}$                              &  $3.75_{-1.29}^{+1.08}$\\
    G6                       &   06:00:07.58       &   -20:08:40.14        &   $21.143\pm0.006$     &  -                    &  $233_{-39}^{+25}$                                                   &   $0.202_{-0.017}^{+0.027}$                       &    $63_{-16}^{+12}$                              &  $2.47_{-1.50}^{+1.05}$\\
    G7                       &   06:00:09.95       &   -20:07:21.31        &   $21.107\pm0.005$     &  -                    &  $97_{-22}^{+45}$                                                    &   $0.098_{-0.038}^{+0.018}$                       &    $22_{-7}^{+13}$                               &  $0.15_{-0.12}^{+0.22}$\\
    G8                       &   06:00:09.85       &   -20:08:03.68        &   $21.266\pm0.004$     &  -                    &  $216_{-13}^{+8}$                                                    &   $0.190_{-0.004}^{+0.007}$                       &    $57_{-10}^{+8}$                               &  $1.95_{-0.71}^{-0.55}$\\
    G9                       &   06:00:05.38       &   -20:08:36.60        &   $21.312\pm0.005$     &  -                    &  $222_{-13}^{+11}$                                                   &   $0.194_{-0.006}^{+0.007}$                       &    $59_{-10}^{+8}$                               &  $2.11_{-0.79}^{+0.63}$\\
    G10                      &   06:00:10.66       &   -20:06:50.65        &   $20.144\pm0.003$     &  -                    &  $220_{-13}^{+14}$                                                   &   $0.193_{-0.009}^{+0.007}$                       &    $59_{-10}^{+9}$                               &  $2.06_{-0.75}^{+0.67}$\\
    G11                      &   06:00:09.58       &   -20:07:00.54        &   $20.271\pm0.004$     &  $0.4439\pm0.0002$    &  $392_{-17}^{+15}$                                                   &   $0.311_{-0.007}^{+0.005}$                       &    $118_{-19}^{+16}$                             &  $13.15_{-4.50}^{+3.75}$\\
    G12                      &   06:00:19.94       &   -20:06:44.71        &   $20.305\pm0.014$     &  -                    &  $293_{-26}^{+23}$                                                   &   $0.244_{-0.015}^{+0.016}$                       &    $83_{-16}^{+13}$                              &  $5.16_{-2.15}^{+1.85}$\\
    G13                      &   06:00:20.40       &   -20:06:39.98        &   $21.029\pm0.014$     &  -                    &  $119_{-12}^{+26}$                                                   &   $0.115_{-0.021}^{+0.009}$                       &    $28_{-6}^{+8}$                                &  $0.28_{-0.13}^{+0.21}$\\
    G14                      &   06:00:13.96       &   -20:07:27.66        &   $21.762\pm0.005$     &  -                    &  $214_{-7}^{+8}$                                                   &   $0.173_{-0.005}^{+0.004}$                       &    $60_{-3}^{+4}$                                &  $2.00_{-0.26}^{+0.28}$\\\hline
    \end{tabular}
    \par
    \begin{justify}
        $^{\mathrm{a}}$ Fixed to the \texttt{SExtractor}-measured values.\\
        $^{\mathrm{b}}$ Scaled to the galaxy luminosity using the relations \eqref{eq:sigma-relation}-\eqref{eq:rcut-relation}.\\
        $^{\mathrm{c}}$ Weight of the galaxy is a free parameter in the SL model (section~\ref{sec:SL_setup}).\\
        $^{\mathrm{d}}$ Core radius is a free parameter in the SL model (section~\ref{sec:SL_setup}).
        
        \par\smallskip\noindent\texttt{Note.} -- \textit{Column~1}: Galaxy designation. -- \textit{Columns~2 and~3}: Right ascension and declination. -- \textit{Column~4}: Observed $r$- or F606W-band magnitude from our GMOS imaging or the RELICS catalog. -- \textit{Column~5}: Spectroscopic redshift from either the GMOS, MUSE or DEIMOS spectroscopy. -- \textit{Column~6}: Velocity dispersion of the dPIE profile. -- \textit{Column~7}: Core radius of the dPIE profile. -- \textit{Column~8}: Truncation radius of the dPIE profile. -- \textit{Column~9}: Total mass of the dPIE profile computed from the model parameters.
    \end{justify}
    \label{tab:cluster_members}
\end{table*}

Due to the fortunate availability of multiple data sets for this cluster, we use several methods to assemble a catalog of cluster members in a hierarchical way.

First, we include all galaxies that have spectroscopic redshifts at the cluster redshift from the MUSE sample of MACS0600 \citep[][]{laporte21b,fujimoto21} and the DEIMOS observations which extend further out into the eastern part of the cluster (see section~\ref{sec:DEIMOS}). The DEIMOS cluster members were selected based on their colors in the UH 2.2\,m telescope imaging and we refer the reader to \citet{fujimoto21} for the MUSE spectroscopic cluster member selection. The combined samples contain 67 spectroscopic cluster members. The distribution of spectroscopic redshifts of the cluster member galaxies and the cluster redshift $z_{\mathrm{d}}=0.432$ (see below) are shown in Fig.~\ref{fig:cluster_members}. While the redshift distribution shown in Fig.~\ref{fig:cluster_members} hints that there may be two distinct over-densities of galaxies at slightly different redshifts, we find the DEIMOS cluster members with $\langle z\rangle=0.445$ to be uniformly distributed across the field. This rather suggests that MACS0600 is a large single cluster with some elongation along the line of sight. We also note that the spectroscopic redshifts of the cluster members slightly depart from the cluster redshift of $z_{\mathrm{d}}=0.46$ as reported from the MACS sample in \citet{ebeling01}. This redshift was solely based on ground-based imaging and spectroscopy of the suspected brightest cluster galaxy (BCG) and therefore relatively uncertain. Averaging over all spectroscopically confirmed cluster members, we find an updated cluster redshift of $z_{\mathrm{d}}=0.432\pm0.007$ which we will use throughout the rest of this study.

Next, we complement this spectroscopic sample with cluster members photometrically selected as lying in the cluster's red sequence in a color-magnitude analysis \citep[e.g.][]{repp18}. We first include the cluster member catalog selected by the red sequence in the RELICS HST catalogs of MACS0600 which was already used in the previous SL models of the cluster \citep[][]{laporte21b,fujimoto21}. These cluster members, along with the spectroscopically selected ones, are then cross-matched with our GMOS catalogs which enables us to calibrate the cluster's red-sequence in $g-r$ color-space. We then select galaxies with \textit{r}-band magnitudes $m_{r}<24$ and within $g-r=\pm0.3$ of the thus defined red sequence as shown in Fig.~\ref{fig:red-sequence}.

Combining these different selections, we obtain a catalog totalling 504 cluster member galaxies across the \textit{Gemini-N} and HST fields-of-view, 67 of which have spectroscopic redshifts. A table of cluster members is available in the online supplementary material of this paper. The new cluster member sample confirms that the eastern structure of the cluster has two BCGs separated by 0.61\arcmin\ (0.21\,Mpc). The RELICS BCG of MACS0600, which we will denominate `South BCG' in accordance with the nomenclature adopted in section~\ref{sec:data}, is separated by 2.88\arcmin\ (0.97\,Mpc) from the cluster's BCG (`BCG1'). The brightest galaxy in the western clump (`West BCG') is separated by 2.15\arcmin\ (0.72\,Mpc) from BCG1 and by 1.21\arcmin\ (0.41\,Mpc) from the South BCG. Interestingly, there is an extremely bright galaxy, the second brightest of our whole sample, in the north-eastern corner of the GMOS field-of-view (about 2.8\arcmin\ or 0.9\,Mpc from BCG1) which also seems to present an overdensity of cluster members around it (see Fig.~\ref{fig:crit-curves}), suggesting an additional sub-clump in the north-east. Since we do not have any SL constraints close to that area (see section~\ref{sec:multiple_images} and Fig.~\ref{fig:multiple-images}), our SL model will not be able to well constrain its mass distribution. The cluster galaxy distribution does however suggest there to be even more extended structures to the cluster than visible in the GMOS imaging though, which we will discuss further in section~\ref{sec:large-scale-structure}. The positions of these BCGs are shown in Fig.~\ref{fig:crit-curves} and their properties summarized in Tab.~\ref{tab:cluster_members}.

\subsection{Multiple image systems} \label{sec:multiple_images}

\begin{figure*}
    \includegraphics[width=\textwidth]{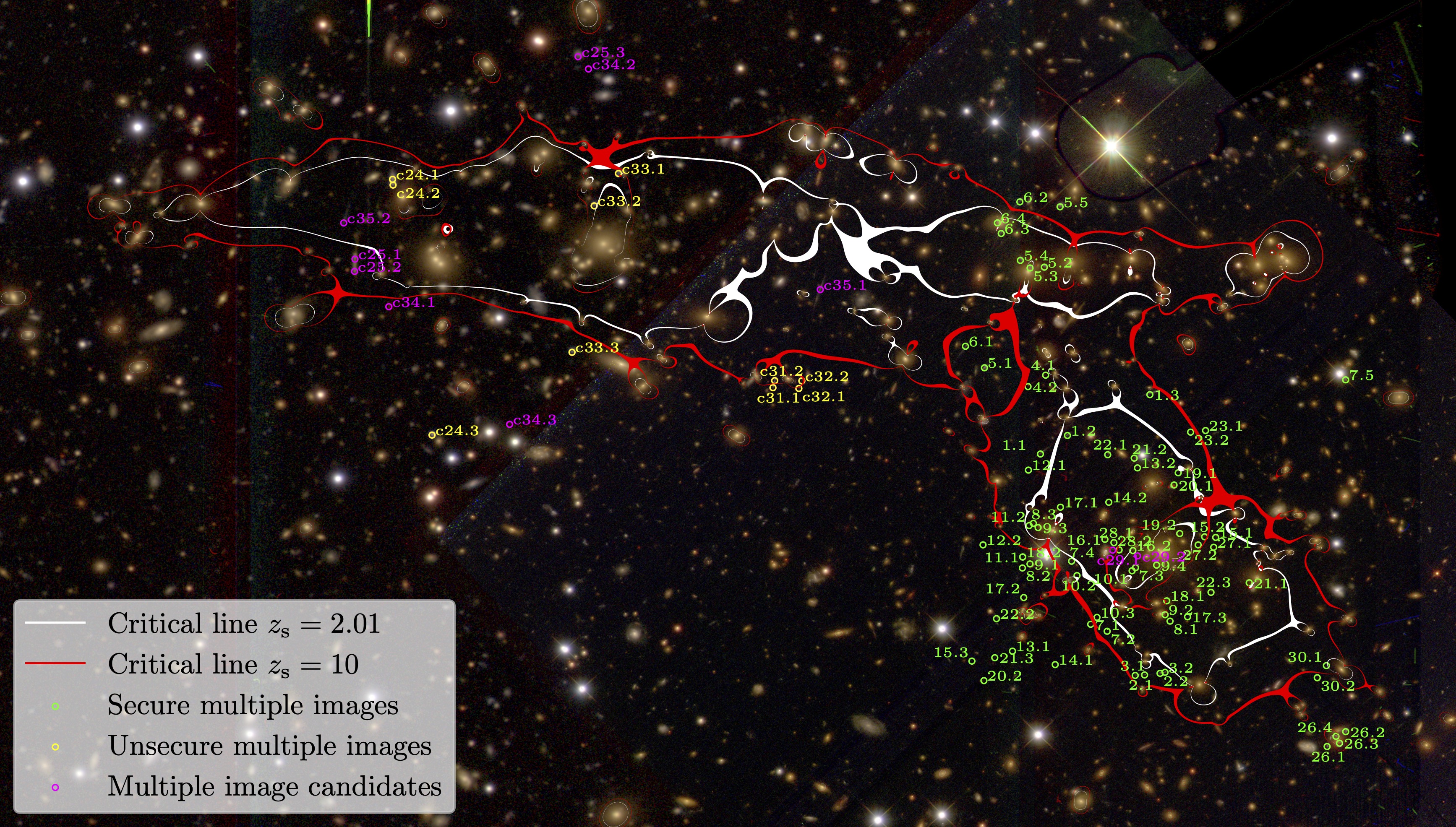}
    \caption{Combined ACS+GMOS composite color image of MACS0600 as in Fig.~\ref{fig:crit-curves}, zoomed-in on the critical area of the cluster core. The critical lines of our SL model for sources at $z_{\mathrm{s}}=2.01$ (corresponding to system 1) and $z_{\mathrm{s}}=10$ are overlaid in white and red respectively. The positions of multiple images are also noted (see also Tab.~\ref{tab:multiple_images}). The \emph{green} images are spectroscopically confirmed images or other photometric images that are considered secure and are used as constraints in the modeling. Images labeled with `c' are candidate systems mostly based on ground-based lower resolution imaging (see section~\ref{sec:gemini}). Candidate images in \emph{magenta} were not used in the modeling but the \emph{yellow} ones were incorporated to anchor the eastern part of the cluster to the more reasonable-looking system candidates in that region. A full-resolution (0.06\arcsec/pix), fully vectorized version of this figure is included in the online supplementary material of this paper.}
    \label{fig:multiple-images}
\end{figure*}

The southern and western clumps of MACS0600 were already studied with the RELICS HST data (see section~\ref{sec:RELICS}) in \citet{laporte21b} and \citet{fujimoto21} which yielded ten multiple image systems, seven of which had spectroscopic redshifts either from the shallow MUSE data available at the time (see section~\ref{sec:multiple_images}) and from GMOS and ALMA \citep{laporte21b,fujimoto21}. Now, thanks to the JWST/NIRCam imaging (see section~\ref{sec:JWST}), we are able to identify 17 additional multiple image systems and one candidate system in the southern clump, 13 of which lie within the new deep MUSE pointing (Fig.~\ref{fig:fields}; see section~\ref{sec:muse}) and thus also have spectroscopic redshifts. Note that we only use spectroscopic redshifts with confidence levels $\geq2$. Unfortunately, the depth and spatial resolution of the \textit{Gemini}/GMOS imaging (section~\ref{sec:gemini}) are not high enough to securely identify multiple image candidates in the eastern part of the cluster. We nevertheless tentatively find two multiple image systems there, which are tentatively supported by our SL model (see section~\ref{sec:SL_result}), and two candidate systems.

All multiple images and candidates are shown in Figs.~\ref{fig:multiple-images} and listed in detail in Tab.~\ref{tab:multiple_images} in appendix~\ref{app:MI-table}. In total, we have identified 91 multiple images and candidates, belonging to 35 sources over the whole cluster field shown in Fig.~\ref{fig:crit-curves}. Systems 1 to 10 were used in the previous SL models of the cluster and system 7 is the $z=6.0719\pm0.0004$ ALMA-detected system \citep{fujimoto21,laporte21b}, quintuply imaged in a hyperbolic-umbilic-like configuration \citep{limousin08,meena20,meena23b,lagattuta23}. The NIRCam imaging now confirms the fifth image of system 7 \citep[image 7.5;][]{fujimoto24} and one of the new NIRCam-identified systems, system 26, is a quadruply-imaged galaxy-galaxy system. Systems 27 and 30 are clearly detected in the NIRCam data but appear neither in the HST nor the MUSE observation and are predicted at high redshift ($z>6$) by our SL model (see section~\ref{sec:SL_result}). In the eastern part of the cluster, our SL model tentatively confirms (i.e. is able to reproduce) system candidates c24 and c33 but not the other candidates found in the GMOS imaging by eye. Finally, system candidates c31 and c32 are identified in the HST imaging, close to its eastern edge (see Fig.~\ref{fig:fields}). We therefore also use them to constrain the newly discovered eastern structures of the cluster due to lack of constraints in that area, even though they are less secure. Note that other candidate images, also marked `c' in Tab.~\ref{tab:multiple_images}, are not used as constraints in the SL model.

\subsection{Model setup} \label{sec:SL_setup}
Since the HST-part of MACS0600 was previously known from RELICS and modeled in \citet{fujimoto21} and \citet{laporte21b}, we follow these models and place two smooth PIEMD \eqref{eq:PIEMD} DM halos, one centered on the south BCG and one centered on the western BCG (see Tab.~\ref{tab:cluster_members} and Fig.~\ref{fig:crit-curves}), and fix their center positions. The extended eastern structure discovered in the GMOS imaging clearly presents two BCGs (Fig.~\ref{fig:crit-curves}), which often hints at there being two cluster-scale DM halos. We therefore place two halos in the main body, initially centered on BCG1 and BCG2. Preliminary SL models prefer the eastern-most halo to be centered north of its BCG. Its center coordinates are therefore left as free parameters whereas the other halo is fixed to be centered on BCG2.

The 504 cluster member galaxies identified in section~\ref{sec:cluster_members} are included in the model as dPIE \eqref{eq:dPIE} profiles. In order to reduce the otherwise gigantic number of free parameters, the cluster member dPIE parameters are scaled to a reference galaxy using luminosity scaling relations:

\begin{align}
    &\sigma_{v}=\sigma_{v,\star}\left(\frac{L}{L_{\star}}\right)^{\lambda} \label{eq:sigma-relation}, \\
    &r_{\mathrm{core}}=r_{\mathrm{core},\star}\left(\frac{L}{L_{\star}}\right)^{\beta} \label{eq:rcore-relation}, \\
    &r_{\mathrm{cut}}=r_{\mathrm{cut},\star}\left(\frac{L}{L_{\star}}\right)^{\alpha} \label{eq:rcut-relation},
\end{align}

\noindent which is a common approach in parametric SL modeling \citep[e.g.][]{halkola06,halkola07,jullo07,monna14,monna15}. We refer the reader to \citet{furtak23c} for more details on the implementation of this method in our code. In our model of MACS0600, we fix the core radius of the reference galaxy to $r_{\mathrm{core},\star}=0.2\,\mathrm{kpc}$ and leave its velocity dispersion $\sigma_{v,\star}$ and truncation radius $r_{\mathrm{cut},\star}$ as free parameters. The reference galaxy luminosity $L_{\star}$ corresponds to an absolute magnitude of $M=-21.04$ (in the \textit{r}-band), which was measured as the Schechter \citep{schechter76} exponential cut-off luminosity of our cluster member sample. The exponents in equations~\eqref{eq:sigma-relation} to~\eqref{eq:rcut-relation} are also free parameters in our model. Finally, it is possible to leave the relative weight of selected galaxies as free parameters if necessary, which we do for 16 cluster galaxies that lie close to multiple images used as constraints (see Tab.~\ref{tab:cluster_members}). The model does not include an external shear component (cf. section~\ref{sec:large-scale-structure}).

The model is constrained by a total of 83 multiple images, belonging to 31 sources, 20 of which have spectroscopic redshifts (see Tab.~\ref{tab:multiple_images}). We assume a positional uncertainty of $\sigma_i=0.5\arcsec$ for each image and give a higher weight to system 7, the $z=6.07$ quintuply-imaged system, by assigning it $\sigma_i=0.25\arcsec$. Spectroscopic multiple image systems have fixed redshifts in the model and the redshifts of photometric systems are left as free parameters. As can be seen in Tab.~\ref{tab:multiple_images}, systems 4, 5, 6, 26 and 28 have relatively well constrained photometric redshifts. We therefore also leave their redshifts as free parameters in the model but limit them to lie within the 95\,\%-range of the photometric redshift. All other redshifts are left free to vary between $z=0.8$ and $z=6$ in the GMOS field-of-view and up to $z=10$ in the HST and JWST fields-of-view (see Fig.~\ref{fig:fields}). We do not use any parity information in this model.

As explained in more detail in section~3.4 in \citet{furtak23c}, the code optimizes the model by minimizing the source plane $\chi^2$, defined as follows:

\begin{equation} \label{eq:SP_Chi2}
    \chi^2=\sum_{i}{\frac{(\Vec{\beta}_i-\Vec{\beta}_0)^2}{\mu_i^2\sigma_i^2}},
\end{equation}

\noindent where $\Vec{\beta}_i$ represents the source position of each multiple image, $\Vec{\beta}_0$ the common modeled barycenter source position of the images belonging to the same system, and $\mu_i$ and $\sigma_i$ the magnification and positional uncertainty of each multiple image. In principle, this method performs as well as a minimization in the lens plane \citep[][]{keeton10} while being significantly faster in terms of computation time. Note that since our model is constrained with the point positions and redshifts of multiple images, and not with surface brightness distributions, we are not concerned about degeneracies between small-scale mass density perturbations and source properties \citep[e.g.][]{wagner22} for the scope of this work. The minimization is performed by first running 35 relatively short Monte-Carlo Markov Chains (MCMC) of $\sim10^3$ steps each to crudely sample the parameters space and estimate the covariance matrix, and then refining the result in a longer MCMC chain with several $10^4$ steps and decreasing ``temperature'' to effectively enable annealing.

\section{Results} \label{sec:result}
The results of our analysis of MACS0600 and its extended eastern structures are presented in the following. In sections~\ref{sec:extended} and~\ref{sec:X-ray_gas}, we summarize the discovery of the eastern structure presented in this work in the optical, X-ray and SZ analyses of the cluster. In section~\ref{sec:SL_result}, we present the results of the SL mass modeling analysis of the extended MACS0600 field.

\subsection{Optical detection of the extended structures} \label{sec:extended}
As introduced in section~\ref{sec:data}, a significant mass outside of the cluster was first suspected due to difficulties in reproducing the strongly lensed multiple images in the HST field-of-view of MACS0600. First mass models for the cluster based on HST/RELICS and the shallow VLT/MUSE data available at the time (see section~\ref{sec:muse}) were presented in the works that first investigated the quintuply imaged $z=6.07$ galaxy, \citet{fujimoto21} and \citet{laporte21b} (see also the discussion in section~\ref{sec:limits_literature}), and required a strong external shear component to properly explain the multiple images. This was perhaps most strongly noted by the model using the \texttt{Light-Traces-Mass} method \citep[\texttt{LTM};][]{zitrin15a}, in which the mass distribution follows the cluster galaxy distribution in the field-of-view such that no intrinsic ellipticity is assigned, making it susceptible to outside massive structures which contribute to the overall lensing signal. While recent works have found that external shear components in SL can possibly arise due to several factors that do not necessarily involve additional cluster structure beyond the field-of-view \citep[][]{lin23,etherington24}, in this particular case some shallow ground-based imaging data were available (see section~\ref{sec:ancillary}) and indeed hinted at possible cluster structures outside HST's field-of-view.

To verify this, we obtained deep optical imaging and spectroscopy with \textit{Gemini-N}/GMOS (section~\ref{sec:gemini}). The GMOS imaging indeed reveals a prominent overdensity of cluster galaxies north-east of the RELICS field-of-view (Fig.~\ref{fig:crit-curves}). Our GMOS spectroscopy, together with the \emph{Keck}/DEIMOS spectroscopy (sections~\ref{sec:gemini} and~\ref{sec:DEIMOS}), confirms the newly discovered eastern structure to lie at the same redshift as the previously observed parts of MACS0600: The redshifts of the two eastern BCGs (BCG1 and BCG2) are $z=0.4403\pm0.0001$ and $z=0.4357\pm0.0002$, respectively (see also Tab.~\ref{tab:cluster_members}). As discussed in detail in section~\ref{sec:cluster_members}, eleven additional cluster member galaxies in the eastern part were spectroscopically confirmed to lie at the cluster redshift as well. 

Using the 67 spectroscopic redshift measurements of cluster members see section~\ref{sec:cluster_members}, we can estimate the velocity dispersions of the cluster substructures. In order to do that, we assign each galaxy with a spectroscopic redshift measurement to one of the sub-structures by determining which BCG it lies closest to and thus obtain tentative velocity dispersions of $\sigma_{v,\mathrm{east}}=(1276\pm203)\,\frac{\mathrm{km}}{\mathrm{s}}$, $\sigma_{v,\mathrm{south}}=(1316\pm146)\,\frac{\mathrm{km}}{\mathrm{s}}$ and $\sigma_{v,\mathrm{west}}=(506\pm132)\,\frac{\mathrm{km}}{\mathrm{s}}$ for the eastern, southern and western sub-structures respectively. The uncertainties are computed with a jackknife sampling method. These results indicate that the eastern structure is probably at least as massive as the previously known southern and western clumps. Note however that these estimates are based only on a handful of spectroscopic redshifts in each clump and can therefore only be seen as very rough first-look estimates.

\subsection{The extended intra-cluster gas} \label{sec:X-ray_gas}

\begin{figure*}
    \centering
    \includegraphics[width=0.498\textwidth, keepaspectratio=true]{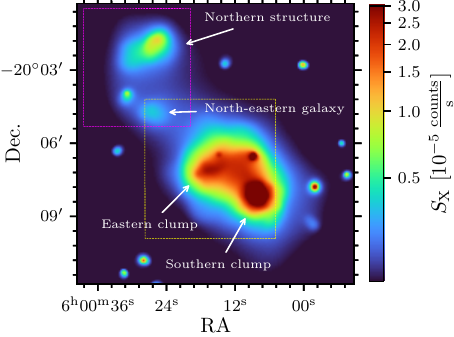}
    \includegraphics[width=0.498\textwidth, keepaspectratio=true]{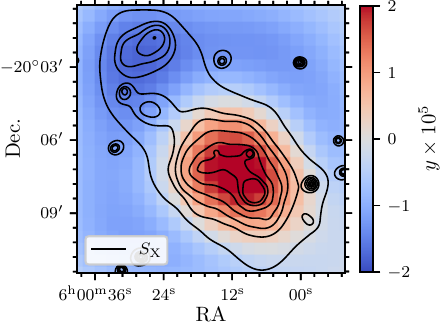}
    \caption{\textit{Left}: X-ray surface brightness map of MACS0600 observed with \textit{XMM-Newton} (section~\ref{sec:ancillary}). The yellow rectangle roughly outlines the area shown in the \textit{Gemini}+HST color image in Fig.~\ref{fig:crit-curves} and the purple rectangle the area shown in Fig.~\ref{fig:norther-structure}. In bulk, the overall intra-cluster gas distribution follows the boot-leg shape of the SL-constrained total mass distribution. The X-ray point-source on the western halo is a bright foreground star. Additional X-ray emitting structures in the far north of the cluster suggest additional massive cluster structures beyond the field-of-view of our optical observations. \textit{Right:} SZ Compton-$y$ map of MACS0600 computed from the ACT and \textit{Planck} data (section~\ref{sec:ancillary}) and smoothed with a $3\times3$\,pixel Gaussian kernel. The black contours represent the X-ray surface brightness in the same steps as shown in Fig.~\ref{fig:crit-curves}. The SZ signal is centered on the barycenter between the eastern, western and southern clumps. The additional northern structures seen in the X-ray map do not appear in the combined ACT and \textit{Planck} SZ map, suggesting that they are below the mass limit of the ACT survey in that sky region, i.e. $M_{500,\mathrm{SZ}}<4\times10^{14}\,\mathrm{M}_{\odot}$.}
    \label{fig:X-ray_gas}
\end{figure*}

We analyze the morphology of the hot ICM gas in MACS0600 through both X-ray observations and the SZ effect.

The X-ray surface brightness map is constructed from the \textit{XMM-Newton} data (section~\ref{sec:ancillary}). We show the cluster's extended X-ray surface brightness in the left-hand panel of Fig.~\ref{fig:X-ray_gas}. The bulk of the X-ray emission is located at the position of the southern DM halo. The overall distribution however follows a boot-leg shape, similar to the cluster member distribution and our SL model mass distribution (Fig.~\ref{fig:crit-curves}), with an overdensity that corresponds to the eastern DM halos. Note that there seems to be an X-ray point source in the western DM halo which concurs spatially with a bright foreground star (see Fig.~\ref{fig:crit-curves}). While the southern component is brightest in X-rays, it is probably slightly less massive than the eastern clump. Its temperature profile decreases from $T_{\mathrm{X}}=(8.3\pm0.5)$\,keV to $T_{\mathrm{X}}=(6.3\pm0.4)$\,keV towards its core whereas the eastern clump has a very high temperature of $T_{\mathrm{X}}>9$\,keV and up to $T_{\mathrm{X}}=(10.0\pm0.6)$\,keV. This suggests that the eastern halo is slightly more massive, whereas the southern structure is in a less advanced stage of merging, since it still has a well-defined cool core. Note that this also concurs with the results of our SL modeling analysis as described in section~\ref{sec:SL_result}.

Interestingly, there is a distinct (i.e. $\sim30\sigma$ detected) extended X-ray structure located far outside even the \textit{Gemini}/GMOS field-of-view, $\sim5\arcmin$ from BCG1 (corresponding to $\sim1.7$\,Mpc at $z_{\mathrm{d}}=0.43$). This strongly hints at further massive cluster structures associated with MACS0600 beyond those discovered using GMOS imaging in this work, even though at present it is not possible to precisely determine if this northern structure lies at the same redshift as the cluster since we do not detect any X-ray emission lines. That being said, we find an overdensity of probable cluster galaxies in that area in the wide-field VIRCAM observations (see section~\ref{sec:ancillary}) which we discuss in detail in section~\ref{sec:large-scale-structure}. In addition, there is another smaller and more diffuse X-ray overdensity that corresponds to the position of the bright north-eastern galaxy seen in the corner of the GMOS imaging (section~\ref{sec:cluster_members} and Fig.~\ref{fig:crit-curves}), which is in the middle between the eastern clump and the additional northern structure seen in the X-ray map. The northern X-ray emitting structure has a temperature of $T_{\mathrm{X}}=(4.5\pm0.3)$\,keV, which indeed confirms it to be a massive cluster-like structure. The smaller diffuse X-ray overdensity on the north-eastern BCG is somewhat cooler with a temperature of $T_{\mathrm{X}}=(3.5\pm0.2)$\,keV. Finally, we note that the X-ray surface-brightness map also shows several point sources, some of them perhaps quasars, located around the cluster.

For examining the cluster's SZ signal, we create a combined SZ-map from the two mm-wave frequency channels from the ACT and \textit{Planck} maps (see section~\ref{sec:ancillary}). We convert both maps to units of the SZ Compton-$y$ parameter and smooth the 150\,GHz map to an effective FWHM of 2.1\arcmin. Then, we compute the inverse-variance weighted average of the two maps. The resulting combined $y$-map is shown in the right-hand panel of Fig.~\ref{fig:X-ray_gas}, overlaid with the X-ray surface-brightness contours. The SZ-signal agrees with the location of the total mass distribution of the cluster, but the spatial resolution is not high enough to show if it is centered on any particular sub-halo. There is nonetheless clearly a contribution from the massive eastern structure, as the signal seems to be centered on the barycenter between the southern, western and eastern clumps. The extended structures to the far north, seen in the X-ray map, do not appear in the SZ map. Given the mass-limit of the ACT survey in this region of the sky \citep{hilton21}, we therefore place a conservative upper limit on the total mass of the northern structures based on the ACT survey limit it that sky region, i.e. $M_{500,\mathrm{SZ}}<4\times10^{14}\,\mathrm{M}_{\odot}$, consistent with the lower X-ray temperatures. The SZ-derived total mass of the cluster from ACT is $M_{500,\mathrm{SZ}}=(9.7\pm1.7)\times10^{14}\,\mathrm{M}_{\odot}$ \citep{hilton21} and from \textit{Planck} is $M_{500,\mathrm{SZ}}=(10.7\pm0.5)\times10^{14}\,\mathrm{M}_{\odot}$ \citep{planck16_xxvii}.

\subsection{SL mass model} \label{sec:SL_result}

\begin{figure}
    \centering
    \includegraphics[width=\columnwidth]{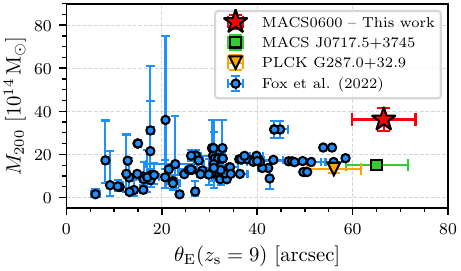}
    \caption{Cluster mass M$_{200}$ in relation to the effective Einstein radius for a source at $z_{\mathrm{s}}=9$. Our results for MACS0600 are shown in red and we also show the equivalent values for the two massive clusters MACS~J0717.5+3745 \citep{zitrin15a} and PLCK~G287.0+32.9 \citep{zitrin17} in green and orange, respectively. For comparison, the \citet{fox22} sample of 74 clusters is shown in blue. With the addition of the eastern structures, MACS0600 clearly ranges among the largest and most massive galaxy clusters known to date.}
    \label{fig:cluster_masses}
\end{figure}

\begin{table*}
    \caption{Median DM halo PIEMD parameters, defined in eqs.~\eqref{eq:PIEMD} and~\eqref{eq:xi} in section~\ref{sec:SL-modeling}, and their $1\sigma$-errors for our SL model of MACS0600.}
    \begin{tabular}{lcccccc}
    \hline
    Halo            &   RA                              &   Dec.                             &   $\epsilon$                  &   $\theta$ [Degrees]   &   $\sigma_v$ [$\frac{\mathrm{km}}{\mathrm{s}}$] &   $r_{\mathrm{core}}$ [kpc]\\\hline
                    &   \multicolumn{6}{c}{\textit{Eastern structure}}\\\hline
    East-1          &   06:00:19.47$_{-0.01}^{+0.02}$   &   -20:06:45.42$_{-1.57}^{+0.86}$   &   $0.819_{-0.003}^{+0.002}$   &  $3.8_{-0.1}^{+0.2}$   &   $742_{-27}^{+28}$                             &  $12_{-3}^{+2}$\\
    East-2          &   06:00:16.81                     &   -20:06:55.65                     &   $0.336_{-0.020}^{+0.012}$   &  $72.7_{-2.9}^{+2.1}$  &   $807_{-18}^{+33}$                             &  $28_{-2}^{+2}$\\\hline
                    &   \multicolumn{6}{c}{\textit{RELICS MACS0600 field-of-view}}\\\hline
    South           &   06:00:08.16                     &   -20:08:08.83                     &   $0.232_{-0.001}^{+0.001}$   &  $-65.6_{-1.3}^{+0.8}$ &   $1203_{-13}^{+18}$                            &  $74_{-3}^{+4}$\\
    West            &   06:00:10.23                     &   -20:07:02.35                     &   $0.531_{-0.009}^{+0.012}$   &  $-0.7_{-0.3}^{+0.4}$  &   $795_{-25}^{30}$                             &  $93_{-15}^{22}$\\\hline
    \end{tabular}
    \par\smallskip
    \begin{justify}
        \texttt{Note.} -- \textit{Column~1}: DM halo designation -- \textit{Columns~2 and~3}: Right ascension and declination. One of the eastern halos and the southern and western halos are fixed to their BCG positions (see section~\ref{sec:SL_setup}). -- \textit{Column~4}: Ellipticity $\epsilon$ defined in equation~\eqref{eq:epsilon}. -- \textit{Column~5}: Position angle, counter-clockwise with respect to the east-west axis. -- \textit{Column~6}: Velocity dispersion of the PIEMD profile. -- \textit{Column~7}: Core radius of the PIEMD profile.
    \end{justify}
    \label{tab:DM-halo_results}
\end{table*}

We construct an SL model for the extended MACS0600 field, including the massive eastern structure. While the shape of the model around the eastern part is based largely on photometric multiple-image candidates identified in the \textit{Gemini} data from the ground and thus requiring further verification, the model suggests that, in line with its optical luminosity density, ICM gas signatures, and velocity dispersion, the eastern part is indeed very massive and contributes significantly the lensing signal. 

The final SL model of the cluster has an average \textit{lens} plane image reproduction error (RMS; root of mean squares) of $\Delta_{\mathrm{RMS}}=0.72\arcsec$ which corresponds to a lens plane $\chi^2=171$. Given the 93 SL constraints and 52 free parameters employed, which result in 41 degrees of freedom, this translates to a reduced $\chi^2$ of $\simeq4.2$. Note that the RMS is mostly driven by the heavily underconstrained eastern part of the cluster. The RMS of multiple images located in the HST- and MUSE observed western and southern parts of the cluster (see Fig.~\ref{fig:multiple-images}) is $\Delta_{\mathrm{RMS,SW}}=0.23\arcsec$. In particular, system 7, the $z=6.072$ quintuply-imaged system further analyzed in \citet{fujimoto24}, is very well reproduced with an average reproduction error $\Delta_{\mathrm{RMS},7}=0.12\arcsec$ between its five images. This means that the well-constrained previously known sub-structures of the cluster are exceptionally well reproduced by our model.

The critical curves for sources at $z_{\mathrm{s}}=2.01$ and $z_{\mathrm{s}}=10$ are shown overlaid over the cluster in Fig.~\ref{fig:crit-curves} and again in Fig.~\ref{fig:multiple-images}. The presented model comprises a total critical area of $A_{\mathrm{crit}}\simeq2.16\,\mathrm{arcmin}^2$ which translates to an effective Einstein radius of $\theta_{\mathrm{E}}=49.7\arcsec\pm5.0\arcsec$, for a source at $z_{\mathrm{s}}=2$, the second largest observed to date after MACS~J0717.5+3745 \citep[$\theta_{\mathrm{E}}=55\arcsec\pm3\arcsec$; e.g.][]{zitrin09b}. The enclosed total mass is then $M(<\theta_{\mathrm{E}})=(4.7\pm0.7)\times10^{14}\,\mathrm{M}_{\odot}$. The southern and western sub-structures, which were previously known from the RELICS observations, have a critical area of $A_{\mathrm{crit,sw}}\simeq0.98\,\mathrm{arcmin}^2$, with an effective Einstein radius of $\theta_{\mathrm{E,sw}}=33.6\arcsec\pm3.4\arcsec$ enclosing $M(<\theta_{\mathrm{E,sw}})=(2.1\pm0.3)\times10^{14}\,\mathrm{M}_{\odot}$ for $z_{\mathrm{s}}=2$. For the same source redshift, the newly discovered eastern sub-structure adds $A_{\mathrm{crit,east}}\simeq1.17\,\mathrm{arcmin}^2$ to the critical area, with $\theta_{\mathrm{E,east}}=36.6\arcsec\pm3.7\arcsec$ enclosing $M(<\theta_{\mathrm{E,east}})=(2.6\pm0.4)\times10^{14}\,\mathrm{M}_{\odot}$. This means that the addition of the eastern structures to the SL model of MACS0600 essentially doubles its critical area. We warrant however that these numbers are still uncertain because our model is poorly constrained in the eastern part due to lack of space-based imaging and secure multiple image identification.

Summing the surface mass density over the whole field of the current SL model (corresponding to the GMOS field-of-view shown e.g. in Figs.~\ref{fig:crit-curves} and~\ref{fig:fields}), we obtain a total projected cluster mass $M_{\mathrm{SL}}$. Given the size of our field-of-view and the typical values for $r_{200}$ of massive galaxy clusters \citep[e.g.][]{merten15}, this is comparable to an $M_{200}$ mass. The result is $M_{\mathrm{SL}}=(3.6\pm0.5)\times10^{15}\,\mathrm{M}_{\odot}$, which makes MACS0600 one of the most massive clusters known to date. It also agrees with the SZ masses derived in section~\ref{sec:X-ray_gas} given typical ratios between $M_{200}$ and $M_{500}$ \citep{merten15}. For comparison, we plot our total cluster mass against the effective Einstein radius (for high-redshift sources in this case) in Fig.~\ref{fig:cluster_masses}, compared to $\sim75$ other massive SL clusters from the literature. As can be seen, MACS0600 is possibly the most extreme case thus far. That being said, we note that while some of the clusters in the \citet{fox22} sample have been found to have associated massive structures outside their field-of-view, e.g. Abell~2744 \citep[e.g.][]{jauzac16,mahler18,furtak23c}, the sample is based on the publicly available SL models which concentrate on small fields-of-view of the cluster centers and would therefore not be sensitive to further extended structures.

The resulting smooth cluster DM halo PIEMD~\eqref{eq:PIEMD} parameters of the model are presented in Tab.~\ref{tab:DM-halo_results}. The most massive single halo is the southern sub-clump and it is also the best constrained since all of the spectroscopic multiple images to date are located in this part of the cluster. The western halo is very elliptical, which also agrees with the projected distribution of cluster galaxies in that area. The two eastern halos, representing the extended sub-structures discovered in this work, are together again as massive as the previously known southern and western structures. These results are further discussed in section~\ref{sec:discussion}.

The resulting dPIE parameters~\eqref{eq:dPIE} of the BCGs and galaxies with a free weight in the model are listed in Tab.~\ref{tab:cluster_members}. For the reference galaxy of luminosity $L_{\star}$, corresponding to $M=-21.04$ (see section~\ref{sec:SL_setup}), our model finds $\sigma_{v,\star}=230_{-9}^{+7}\,\frac{\mathrm{km}}{\mathrm{s}}$ and $r_{\mathrm{cut},\star}=62_{-10}^{+8}\,\mathrm{kpc}$. The resulting luminosity scaling relation exponents from equations~\eqref{eq:sigma-relation} to~\eqref{eq:rcut-relation} are $\lambda=0.488_{-0.003}^{+0.004}$, $\beta=0.406_{-0.005}^{+0.003}$ and $\alpha=0.592_{-0.003}^{+0.003}$. This roughly agrees with typical values for $\lambda$, for example, measured with internal kinematic stellar modeling of cluster galaxies \citep[$\simeq0.4$, e.g.][]{bergamini19,bergamini21,bergamini23a}.

Finally, the redshifts predicted by the model for the systems whose redshifts were left as free parameters (see section~\ref{sec:SL_setup}) are listed in Tab.~\ref{tab:multiple_images} in appendix~\ref{app:MI-table}. We in particular note that the two new systems 27 and 30, which were identified in the JWST/NIRCam imaging, are predicted at relatively high redshifts of $z_{\mathrm{model}}=7.0_{-0.2}^{+0.1}$ and $z_{\mathrm{model}}=6.7_{-0.3}^{+0.2}$, respectively. For system 27, this is consistent with a non-detection in even the deep MUSE data available since the instrument can only detect the Lyman-$\alpha$ line out to $z=6.7$ and the HST/WFC3 data are probably too shallow (0.5\,orbits) to detect this system. System 30 on the other hand lies outside of both the MUSE and WFC3 fields-of-view (see Fig.~\ref{fig:fields}) and would therefore not have been detected previously.

\section{Discussion} \label{sec:discussion}
The optical, ICM, and SL analyses described in section~\ref{sec:result} reveal a previously un-noted massive structure associated with MACS0600, making it one of most massive galaxy cluster studied to date. The model also suggests that it is possibly the cluster with the second-largest Einstein radius of ($\theta_{\mathrm{E}}\simeq49\arcsec$, for a source at $z_{\mathrm{s}}=2$), which could in turn become the largest Einstein radius cluster for high-redshift sources at $z\sim9-10$ (see Fig. \ref{fig:cluster_masses}). In this section, we place these results into context and discuss them with regard to previous SL models and SL modeling systematics (section~\ref{sec:limits_literature}), the large-scale structure around the cluster (section~\ref{sec:large-scale-structure}), and the prospects for future high-redshift surveys using the SL magnification of the cluster (section~\ref{sec:high-z_potential}).

\subsection{Limits of the SL model and comparison to the literature} \label{sec:limits_literature}

\begin{figure}
    \centering
    \includegraphics[width=\columnwidth]{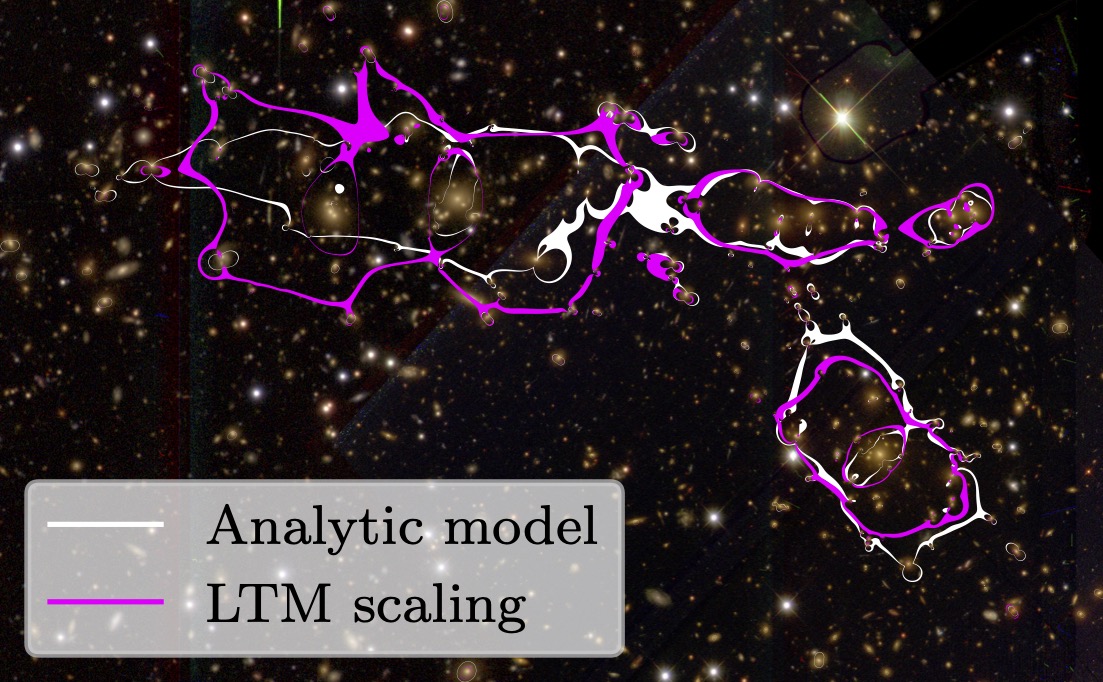}
    \caption{Critical lines for a source at $z_{\mathrm{s}}=2.01$ of our analytic SL model presented in section~\ref{sec:SL_result} (white) and of the \texttt{LTM} model run with only the western multiple images (magenta). The cutout shown here is the same field-of-view as Fig.~\ref{fig:multiple-images}. While no multiple images in the eastern part were used for that model, the \texttt{LTM} model clearly predicts a supercritical massive SL structure that concurs with our eastern halos. A full-resolution (0.06\arcsec/pix), fully vectorized version of this figure is included in the online supplementary material of this paper.}
    \label{fig:LTM-model}
\end{figure}

Previous SL models of MACS0600 were published in \citet{laporte21b} and \citet{fujimoto21} and include two parametric models with \texttt{GLAFIC} \citep{oguri10b} and \texttt{lenstool} \citep{kneib96,jullo07,jullo09}, and an \texttt{LTM} model. Another \texttt{lenstool} parametric model can be found in the public RELICS repository \citep{fox22}. These models were typically based on 10 multiple image systems (systems 1-10 in Tab.~\ref{tab:multiple_images}) known from the RELICS HST imaging, with 7 spectroscopic redshifts from ALMA, GMOS and the shallow MUSE data available at the time (see section~\ref{sec:muse}), and limited to the southern and western halos covered by HST. The public RELICS \texttt{lenstool} model was constrained with 6 multiple image systems and 5 spectroscopic redshifts. Similar to our model, the previous models assumed two or three smooth DM halos to represent the western and southern halos. The model we present here covers the newly discovered eastern structures, and uses additional multiple images and spectroscopic redshifts now available in the original HST region (see Fig.~\ref{fig:multiple-images}).

The main limit of our extended SL analysis is the lack of high-resolution imaging around the eastern halos. While our results for the western and southern halos broadly concur with the literature models mentioned above, the two eastern halos are largely under-constrained, containing only four multiple-image candidate systems (see Fig.~\ref{fig:multiple-images}) and none of them with spectroscopic redshifts. This drives the total average RMS of our SL model $\Delta_{\mathrm{RMS}}=0.72\arcsec$ since the lens-plane RMS is known to often become larger when less spectroscopic multiple image systems are used \citep[e.g.][]{johnson16}. Indeed, the RMS of our model in the western and southern regions of the cluster that are covered by HST and MUSE, is much better: $\Delta_{\mathrm{RMS,SW}}=0.23\arcsec$. The lack of SL constraints in the eastern area might also explain why the first eastern halo strays far from its main BCG, and its somewhat extreme ellipticity (Tab.~\ref{tab:DM-halo_results}). 
Despite this uncertainty, in terms of total SL mass the eastern substructure is predicted to be as or more massive than the previously known parts of MACS0600, essentially doubling the cluster's total mass as shown in section~\ref{sec:SL_result}. This is consistent also with the X-ray temperatures, which suggest that the eastern structure is possibly the more massive one.

To further assess the size of the eastern clumps and the robustness of GMOS-detected multiple image systems in the eastern halos and their affect on the results, we exploit the predictive power of the \texttt{LTM} method \citep{zitrin15a,carrasco20}. We run an \texttt{LTM} SL model using the secure multiple image systems in the western and southern halos and the full cluster member catalog assembled in section~\ref{sec:cluster_members}, but not using any of the eastern multiple images that were only identified in the ground based data and thus remain uncertain. By using the light distribution of cluster members in the eastern part of the cluster alone, the \texttt{LTM} model ($\Delta_{\mathrm{RMS}}=2.1\arcsec$) indeed predicts there to be a significant supercritical SL cluster structure around the two eastern BCGs where we placed the eastern halos in our analytic model (sections~\ref{sec:SL_setup} and~\ref{sec:SL_result}), and with similar symmetry, as can be seen in Fig.~\ref{fig:LTM-model}. Finally, for further verification, we re-run our parametric model with the same configuration but without including the eastern multiple images. This model also achieves an RMS of $\Delta_{\mathrm{RMS}}=0.2\arcsec$. While the individual parameters of the eastern halos are very different than those of our results presented in section~\ref{sec:SL_result} and Tab.~\ref{tab:DM-halo_results}, on account of being completely unconstrained, the overall properties such as the effective Einstein radius, enclosed mass and total cluster mass only vary marginally and agree within $1\sigma$. The results in the southern and western halos from this model also agree with our total model, further confirming that this part of the cluster is well constrained.

In order to better constrain the SL mass profile of the eastern structures, high-resolution space-based observations with HST and JWST to detect more and more robust multiple images and arcs in the eastern part of the cluster would be required. Additional spectroscopic coverage of both the western and eastern halos would also be needed to derive spectroscopic redshifts of new multiple image systems and optimally constrain the total mass distribution of the cluster with SL mass modeling. 

\subsection{Further surrounding large-scale structure} \label{sec:large-scale-structure}

\begin{figure}
    \centering
    \includegraphics[width=0.5\textwidth, keepaspectratio=true]{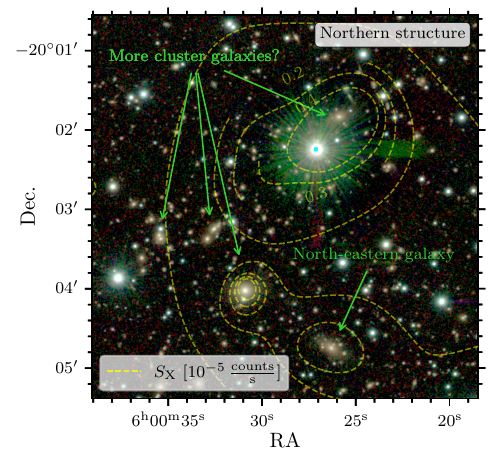}
    \caption{A $4.8\arcmin\times4.8\arcmin$ cutout of the VIRCAM color-composite image used in Fig.~\ref{fig:fields} (blue: \textit{Y}-band, green: \textit{J}-band, red: \textit{Ks}-band) showing the northern X-ray emission structure (see left-hand panel of Fig.~\ref{fig:X-ray_gas}). The X-ray surface-brightness contours are overlaid in yellow as in Fig.~\ref{fig:crit-curves}. The X-ray emission in this region concurs with an overdensity of galaxies that match the colors of cluster members, suggesting that it likely lies at the same redshift as the cluster.}
    \label{fig:norther-structure}
\end{figure}

\begin{figure*}
    \centering
    \includegraphics[width=\textwidth]{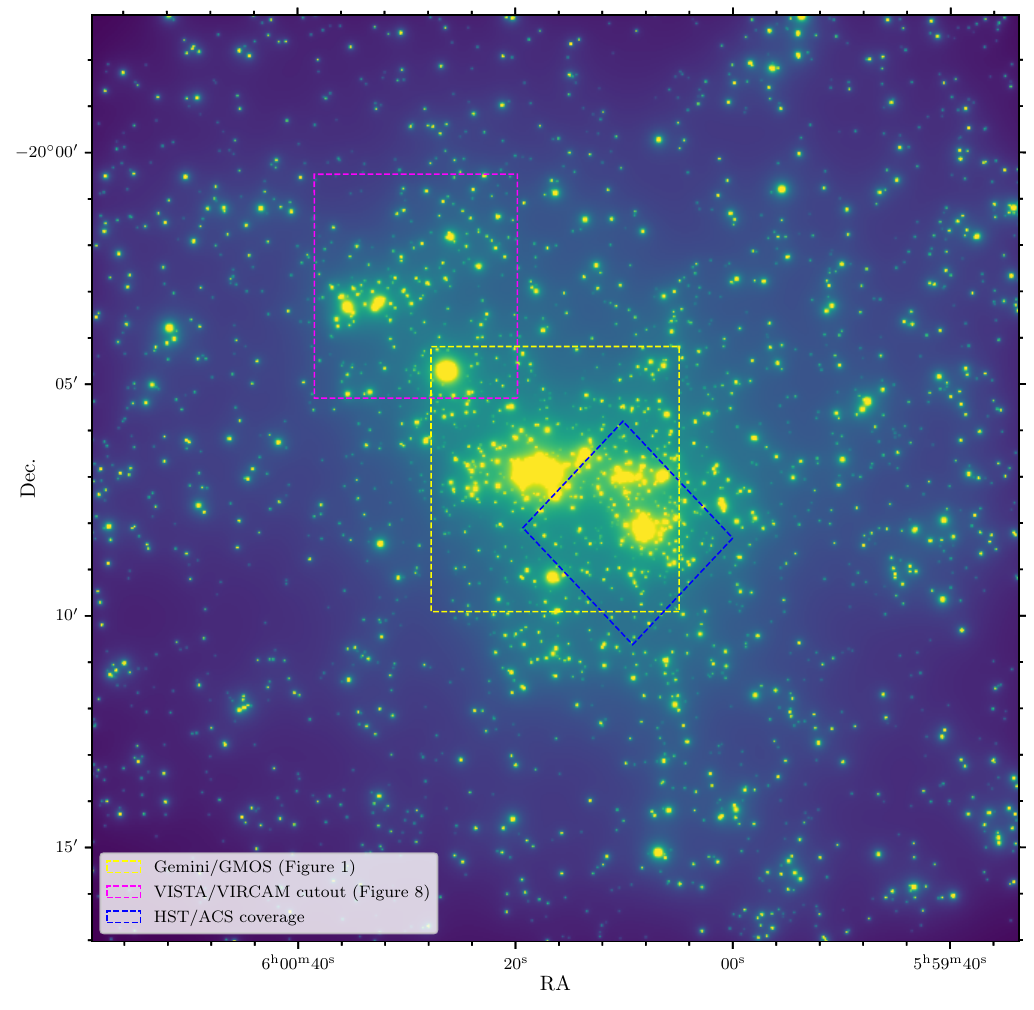}
    \caption{VIRCAM \textit{Y}-band luminosity density of cluster member galaxies in a wide $20\arcmin\times20\arcmin$ area around MACS0600. At $z_{\mathrm{d}}=0.43$, one arcminute corresponds to 0.34\,Mpc. While some 60 galaxies were verified spectroscopically (see section~\ref{sec:cluster_members}), most of the cluster members were selected photometrically, some only in the VIRCAM imaging data with poor color range, and should thus be referred to as mere candidates. The yellow square represents the area shown in Fig.~\ref{fig:crit-curves}, the magenta square the area shown in Fig.~\ref{fig:norther-structure} and the blue square the previously known part of the cluster with RELICS HST coverage. The cluster member luminosity distribution presents an overdensity at roughly the same location as the northern X-ray structures shown in Figs.~\ref{fig:X-ray_gas} and~\ref{fig:norther-structure}, indicating that it indeed most likely lies at the same redshift and is associated with the cluster. Elongations of the luminosity density in the north-east to south-west, and tentatively towards the north-west direction, indicate filamentary structures with the cluster embedded in a node of the cosmic web.}
    \label{fig:luminosity-density}
\end{figure*}

As shown in Figs.~\ref{fig:crit-curves} and~\ref{fig:X-ray_gas}, our analysis finds MACS0600 to be a very large and complex merging galaxy cluster structure, mostly extending along the north-east to south-west direction. While the X-ray analysis described in section~\ref{sec:X-ray_gas} clearly detects another large cluster-scale emission feature in the north, the absence of X-ray emission lines does not enable us to confirm that it lies at the same redshift of $z_{\mathrm{d}}=0.43$ of the cluster.

We instead use the wide-field VIRCAM observations (see section~\ref{sec:ancillary}) to search for cluster galaxies in or around this X-ray overdensity. A VIRCAM color-composite image of the northern X-ray emission feature is shown in Fig.~\ref{fig:norther-structure}. The VIRCAM image clearly shows a number of large elliptical galaxies with similar colors as our cluster members in the vicinity of the X-ray emission. For a more quantitative assessment, we run \texttt{SExtractor} to detect sources and measure photometry in the three VIRCAM bands and then select potential cluster galaxies in color-magnitude and color-color space by calibrating our selection to the VIRCAM colors of our cluster member catalog assembled in section~\ref{sec:cluster_members}. Given the shallow depth of the VIRCAM observations ($23.5$\,magnitudes at $5\sigma$), we select galaxies brighter than 23\,magnitudes for this analysis. While NIR data are not optimal to select cluster galaxies at $z_{\mathrm{d}}=0.43$, since the 4000\,\AA-break falls in optical wavelengths, from the ancillary data listed in section~\ref{sec:ancillary}, the VIRCAM imaging achieve sufficient resolution and depth in a wide field around the cluster. Moreover, our selection recovers all of the spectroscopic and optical photometric cluster members that are bright enough to be robustly detected. This selection indeed confirms most of the galaxies seen in Fig.~\ref{fig:norther-structure} to have the same VIRCAM colors as our cluster members. Furthermore, the resulting luminosity surface density of cluster members, as shown in Fig.~\ref{fig:luminosity-density}, clearly presents an overdensity in the north-east, outside the field of view of our optical imaging. Overall, the luminosity density distribution is complex, mostly elongated in the north-east to south-west direction and with perhaps another branch in the north-western direction. This suggests that MACS0600 and its extended structures form a node of the cosmic web with filaments branching out. Similar structures have been found around other very massive galaxy clusters, most prominently MACS~J0717.5+3745 \citep[e.g.][]{jauzac12,jauzac18,durret16,ellien19}.

We also note that the additional northern structure lies at $\sim5\arcmin$ ($\sim1.7$\,Mpc) from BCG1, which means that it could be targeted in the parallel field of HST if the primary field is aimed at the eastern halos. Future high-resolution HST imaging programs of the eastern extension of MACS0600 could therefore also target the northern structures in the same visit, potentially yielding two large SL fields simultaneously at minimal cost.

As was shown by e.g. \citet{acebron17} and \citet{mahler18}, massive structures far from the cluster center can also significantly affect the SL signal. Thus, the tentative northern structure  may affect the multiple image formation in the cluster core to some extent. In that sense we note, however, that an alternative version of our model, run with an external shear component included, did not significantly impact the results or the multiple image reproduction, suggesting that the contribution from that tentative northern clump on the main SL regime would probably be minor. This may however also be a result of the low number and generally high uncertainty of the eastern multiple image systems which make it impossible to determine if an external shear is currently required since the model is already underconstrained in that area.

\subsection{Prospects for high-z galaxy observations} \label{sec:high-z_potential}

\begin{figure}
    \centering
    \includegraphics[width=\columnwidth]{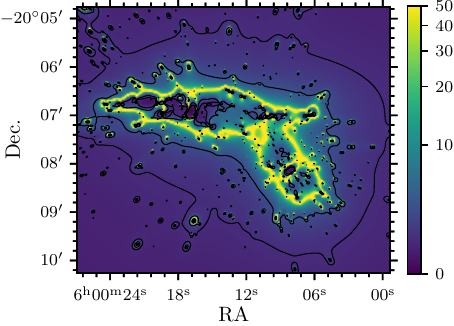}
    \caption{Magnification map of our SL model of MACS0600 for a source at $z_{\mathrm{s}}=10$. The black contours represent magnifications of $\mu=2$ and $\mu=4$. The combination of the substructures in MACS0600 form a large area of high magnification, making the cluster particularly well suited to probe the high-redshift Universe as a cosmic telescope.}
    \label{fig:magnification}
\end{figure}

\begin{figure}
    \centering
    \includegraphics[width=\columnwidth]{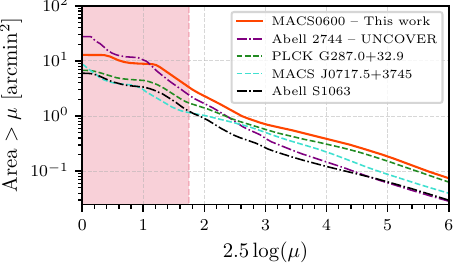}
    \caption{Cumulative source plane area at $z_{\mathrm{s}}=10$ as a function of magnification. Our model of MACS0600 is shown in red and we compare it to models of other very large SL clusters, namely the UNCOVER \citet{furtak23c} model of Abell~2744, the \citet{zitrin17} \texttt{LTM} model of PLCK~G287.0+32.9 and \texttt{LTM} models of MACS~J0717.5+3745 and Abell~S1063 \citep[][]{zitrin15a} for reference. While the low-magnification end ($\mu\lesssim5$, red shaded area) depends only on the size of the observed field, the high-magnification fraction the source plane area is largely independent of that. In this regime, MACS0600 has among the largest source plane areas according to our model.}
    \label{fig:SP_area}
\end{figure}

With its exceptionally large critical area, of $2.16\,\mathrm{arcmin}^2$ at $z_{\mathrm{s}}=2$ (see section~\ref{sec:SL_result}) and $3.91\,\mathrm{arcmin}^2$ at $z_{\mathrm{s}}=10$, MACS0600 represents an excellent field for future studies of the high-redshift Universe using its SL magnification. Indeed, SL cluster fields have been proven invaluable to the study of faint and low-mass objects at high redshifts both with HST \citep[e.g.][]{bouwens17,ishigaki18,atek18,furtak21,strait20,strait21} and JWST \citep[e.g.][]{adams23,atek23a,atek24a,harikane23,bradley23,bouwens23}.

To illustrate that MACS0600 is a promising field for deep high-redshift observations, we show the $z_{\mathrm{s}}=10$ magnification map and cumulated source-plane area as a function of magnification in Fig.~\ref{fig:magnification}, and compare the latter to several other large cluster lenses. The four halos probed in our SL model form a contiguous sheet of high magnification and the high-magnification ($\mu\gtrsim5$) area in the source plane clearly exceeds even the multiple SL cores of Abell~2744 observed with the JWST \textit{Ultradeep NIRSpec and NIRCam ObserVations before the Epoch of Reionization} survey \citep[UNCOVER;][]{bezanson24}, the largest SL field observed to date with JWST. In general, as can be seen in Fig.~\ref{fig:SP_area}, the model of MACS0600 reveals it to present cumulated high-magnification $z_{\mathrm{s}}=10$ source-plane areas of the same order of magnitude as the most massive known clusters, e.g. MACS~J0717.5+3745 \citep[e.g.][]{zitrin09b} and PLCK~G287.0+32.9 \citep{zitrin17,daddona24}, and much larger than other more typical clusters such as Abell~S1063 \citep[e.g.][]{zitrin15a}. While the magnification areas are also strongly dependent on the lens modeling method used \citep[e.g.][]{bouwens17,atek18}, this is nevertheless indicative that MACS0600 ranges among the most efficient cluster lenses in terms of high-magnification area. Note that in the low-magnification regime, the cumulated source plane area is not dominated by the SL region but by the total survey area, which is much larger \citep[i.e. $\sim35\,\mathrm{arcmin}^2$;][]{atek23b} in the UNCOVER survey than in this work. Indeed, our conclusions are based on the current model which is in part based on candidate multiple images requiring verification. Nevertheless, even if the currently implied large critical area of MACS0600 is found to be somewhat overestimated, the existence of the massive eastern structure necessarily boosts the area of high magnifications.

We also note that the southern halo of MACS0600 alone contains a spectacular quintuply imaged and ALMA-detected $z=6$ disc galaxy \citep{fujimoto24}, and likely two more multiply-imaged $z>6$ objects (see section~\ref{sec:SL_result} and Tab.~\ref{app:MI-table}). The RELICS HST data of the cluster yielded a total of 8 $z\gtrsim6$ objects \citep[][]{salmon20}. This all suggests that MACS0600 covers a rich volume of the high-redshift Universe and that many more high-redshift objects may lie behind the eastern halos and be magnified by them. Finally, with the northern potential SL structure, discovered in the X-ray emission (see sections~\ref{sec:X-ray_gas} and~\ref{sec:large-scale-structure}), in range of the HST parallel field, the extended structures of the cluster represent an excellent target for the next generations of deep HST and JWST surveys.

\section{Conclusion} \label{sec:conculsion}
We present a new optical, X-ray, SZ, and lensing analysis of the massive RELICS galaxy cluster MACS0600 in which we discover and characterize an additional massive cluster structure.

In summary, our main results are the following:

\begin{itemize}
    \item We discover a massive, bi-modal extended cluster structure to the north-east of the RELICS HST field-of-view of the galaxy cluster MACS0600 and confirm it to lie at the same redshift. Using spectroscopic redshifts of 67 cluster galaxies, we find an updated cluster redshift of $z_{\mathrm{d}}=0.432\pm0.007$.
    \item We perform an SL analysis of the extended cluster and identify 18 new multiple image systems and candidates in new JWST/NIRCam imaging of the southern part of the cluster and 5 tentative multiple system candidates in the GMOS imaging of the eastern part. New ultra-deep VLT/MUSE observations of the southern sub-clump allow us to spectroscopically confirm 13 multiple image systems. 
    \item The newly discovered extended structure essentially doubles the cluster's critical area, to $A_{\mathrm{crit}}\simeq2.16\,\mathrm{arcmin}^2$ for a source at $z_{\mathrm{s}}=2$, resulting in an effective Einstein radius of $\theta_{\mathrm{E}}=49.7\arcsec\pm5.0\arcsec$ which encloses a total mass of $M(<\theta_{\mathrm{E}})=(4.7\pm0.7)\times10^{14}\,\mathrm{M}_{\odot}$. If confirmed with further high-resolution imaging of the eastern halos of the cluster, this would make MACS0600 the second largest SL cluster observed to date after MACS~J0717.5+3745, with an exceptionally large high-magnification area. The total projected mass of the cluster in the modeled field-of-view is $M_{\mathrm{SL}}=(3.6\pm0.5)\times10^{15}\,\mathrm{M}_{\odot}$, according to the current SL model.
    \item The \textit{XMM-Newton} $0.5-7$\,keV data revealed the cluster's hot X-ray gas to follow the same overall bootleg shape of the total mass distribution. The eastern X-ray clump has a high temperature of $T_{\mathrm{X}}>9$\,keV whereas the southern X-ray clump is cooler ($T_{\mathrm{X}}\simeq6-8$\,keV) and has a decreasing temperature gradient towards its core. This suggests that the eastern part is more massive, although less bright in X-rays, corroborating the findings of the SL analysis.
    \item The cluster is also detected in ACT+\textit{Planck} SZ maps. Although these are of low resolution, the SZ signal is centered roughly on the barycenter of the three main (eastern, southern and western) clumps of the cluster. The SZ-derived total mass of the cluster is $M_{500,\mathrm{SZ}}=(9.7\pm1.7)\times10^{14}\,\mathrm{M}_{\odot}$.
    \item Further structures tentatively associated with MACS0600 are also detected $5\arcmin$ ($1.7$\,Mpc) north-east of the cluster in X-rays and wide-field VISTA/VIRCAM NIR imaging. Additionally, the luminosity distribution of surrounding galaxies shows various filamentary-like structures around the cluster.
\end{itemize}

While our SL model of the eastern structures detected in MACS0600 is corroborated by both the X-ray analysis and an alternative \texttt{LTM} model, it nevertheless remains uncertain due to the lack of SL constraints (and spectroscopic redshifts) in that area. Future models will therefore require high-resolution space-based imaging observations with HST and JWST, and additional supporting spectroscopy, to robustly detect multiple images in that area and properly constrain the cluster's mass distribution. In addition, weak lensing analyses in a wide field around the cluster and its associated structures in the far north, which can be done with wide-field ground-based imaging, as well as including the far northern structures in future SL analyses, would further help to constrain the total mass distribution of MACS0600. With its exceptionally large SL area, and the potential additional SL cluster structure in the north observable in parallel, the Anglerfish cluster MACS0600 represents a prime field for the next generation of ultra-deep HST and JWST surveys to probe the high-redshift Universe through its SL magnification. With their wide fields-of-view, upcoming space-based imaging missions such as e.g. \textit{Euclid} and \textit{Roman} will routinely enable the detections of massive associated cluster structures around SL clusters \citep[e.g.][]{atek24b}.

\section*{Acknowledgements}
We warmly thank the anonymous referee for their very useful comments which greatly helped to improve the paper. The BGU group acknowledges support by Grant No. 2020750 from the United States-Israel Binational Science Foundation (BSF) and Grant No. 2109066 from the United States National Science Foundation (NSF); by the Ministry of Science \& Technology, Israel; and by the Israel Science Foundation Grant No. 864/23. J.S. and H.L. acknowledge support from NASA/ADAP 80NSSC19K1018. M.L. acknowledges the Centre National de la Recherche Scientifique (CNRS) and the Centre National des Etudes Spatiales (CNES) for their support. K.K. acknowledges the support by JSPS KAKENHI Grant Numbers JP17H06130, JP22H04939, and JP23K20035,
and the NAOJ ALMA Scientific Research Grant Number 2017-06B. This research was supported by the \textit{International Space Science Institute} (ISSI) in Bern, through the ISSI International Team project \#476 (``Cluster Physics From Space To Reveal Dark Matter'').

This work makes use of observations obtained at the international \textit{Gemini} Observatory, a program of NSF’s NOIRLab, which is managed by the Association of Universities for Research in Astronomy (AURA) under a cooperative agreement with the NSF. The Gemini Observatory partnership comprises: the NSF (United States), National Research Council (Canada), Agencia Nacional de Investigaci\'{o}n y Desarrollo (Chile), Ministerio de Ciencia, Tecnolog\'{i}a e Innovaci\'{o}n (Argentina), Minist\'{e}rio da Ci\^{e}ncia, Tecnologia, Inova\c{c}\~{o}es e Comunica\c{c}\~{o}es (Brazil), and Korea Astronomy and Space Science Institute (Republic of Korea). Some of the data presented herein were obtained at \textit{Keck} Observatory, which is a private 501(c)3 non-profit organization operated as a scientific partnership among the California Institute of Technology, the University of California, and NASA. The Observatory was made possible by the generous financial support of the W. M. Keck Foundation. We note that the \textit{Keck}, UH, and \textit{Gemini-N} telescopes are located on Maunakea. We are grateful for the privilege of observing the Universe from a place that is unique in both its astronomical quality and its cultural significance. This work is further based on observations obtained with the NASA/ESA \textit{Hubble Space Telescope} (HST) and the NASA/ESA/CSA JWST, retrieved from the \texttt{Mikulski Archive for Space Telescopes} (\texttt{MAST}) at the \textit{Space Telescope Science Institute} (STScI). STScI is operated by the Association of Universities for Research in Astronomy, Inc. under NASA contract NAS 5-26555. In addition, this study is based on observations obtained with \textit{XMM-Newton}, an ESA science mission with instruments and contributions directly funded by the ESA member states and NASA. Finally, this work is also based on observations made with ESO Telescopes at the La Silla Paranal Observatory obtained from the ESO Science Archive Facility.

This research made use of \texttt{Astropy},\footnote{\url{http://www.astropy.org}} a community-developed core Python package for Astronomy \citep{astropy13,astropy18} as well as the packages \texttt{NumPy} \citep{vanderwalt11}, \texttt{SciPy} \citep{virtanen20} and \texttt{Matplotlib} \citep{hunter07}.


\section*{Data Availability}
The RELICS HST data and catalogs on which this work is based are publicly available on the \texttt{MAST} archive in the RELICS \citep[][]{coe19} repository for MACS0600\footnote{\url{https://archive.stsci.edu/missions/hlsp/relics/rxc0600-20/}}. The HAWK-I and VIRCAM data are publicly available on the ESO Science Archive
\footnote{\url{http://archive.eso.org/scienceportal/home}} under program IDs 0103.A-0871(B) and 198.A-2008(E) respectively. The reduced MUSE data cubes are also available on the ESO Science Archive under program IDs 0100.A-0792(A) and 109.22VV.001(A), while high-level spectroscopy products are available on the \citet[][]{richard21} Atlas of MUSE observations towards massive lensing clusters website\footnote{\url{https://cral-perso.univ-lyon1.fr/labo/perso/johan.richard/MUSE_data_release/}}. The JWST/NIRCam observations are publicly available on the \texttt{MAST} archive under program ID GO-1567. The X-ray \textit{XMM-Newton} data are publicly available on the \textit{XMM-Newton} Science Archive\footnote{\url{http://nxsa.esac.esa.int/nxsa-web/\#home}} under observation IDs~0650381401 and~0827050601. The ACT maps \citep{naess20} are publicly available on the \texttt{LAMBDA} server\footnote{\url{https://lambda.gsfc.nasa.gov/product/act/actpol_prod_table.cfm}}. Additional data and reduced data products, i.e. the \textit{Gemini-N}/GMOS and \textit{Keck}/DEIMOS imaging and spectroscopy, will be shared by the authors on request.


\bibliographystyle{mnras}
\bibliography{references} 



\appendix

\section{Multiple-image table} \label{app:MI-table}

\begin{table*}
    \caption{Multiple images and candidates identified in the Anglerfish cluster.}
    \begin{tabular}{lcccccr}
    \hline
    ID      &   RA              &   Dec.            &   $z_{\mathrm{spec}}$         &   $z_{\mathrm{phot}}$~[95\,\%-range]  &   $z_{\mathrm{model}}$~[95\,\%-range] &   Remarks\\\hline
    1.1     &   06:00:10.038    &   -20:07:45.283   &   $2.0032\pm0.0001$           &   -                                   &   -                                   &   \\
    1.2     &   06:00:09.619    &   -20:07:40.883   &   $2.0033\pm0.0002$           &   -                                   &   -                                   &   \\
    1.3     &   06:00:08.341    &   -20:07:31.183   &   -                           &   -                                   &   -                                   &   \\\hline
    2.1     &   06:00:08.422    &   -20:08:37.754   &   $2.7699\pm0.0002$           &   -                                   &   -                                   &   \\
    2.2     &   06:00:08.182    &   -20:08:37.379   &   $2.7734\pm0.0002$           &   -                                   &   -                                   &   \\\hline
    3.1     &   06:00:08.567    &   -20:08:37.879   &   $2.7699\pm0.0002$           &   -                                   &   -                                   &   \\
    3.2     &   06:00:08.105    &   -20:08:37.129   &   $2.7734\pm0.0002$           &   -                                   &   -                                   &   \\\hline
    4.1     &   06:00:09.957    &   -20:07:26.558   &   -                           &   -                                   &   4.61~$[4.59,4.63]$                  &   \\
    4.2     &   06:00:10.230    &   -20:07:29.278   &   -                           &   4.8~$[4.7,5.0]$                     &   "                                   &   \\\hline
    5.1     &   06:00:10.906    &   -20:07:24.800   &   -                           &   1.7~$[1.6,1.8]$                     &   1.84~$[1.77,1.87]$                  &   \\
    5.2     &   06:00:09.978    &   -20:07:00.900   &   -                           &   -                                   &   "                                   &   Close to West BCG.\\
    5.3     &   06:00:10.199    &   -20:07:01.100   &   -                           &   -                                   &   "                                   &   "\\
    5.4     &   06:00:10.349    &   -20:06:59.300   &   -                           &   1.1~$[1.0,1.2]$                     &   "                                   &   "\\
    5.5     &   06:00:09.734    &   -20:06:46.570   &   -                           &   0.2~$[0.1,2.2]$                     &   "                                   &   \\\hline
    6.1     &   06:00:11.200    &   -20:07:19.700   &   -                           &   1.6~$[1.3,1.7]$                     &   1.66~$[1.63,1.69]$                  &   Tentative $z_{\mathrm{spec}}\simeq1.5774$ from GMOS.\\
    6.2     &   06:00:10.357    &   -20:06:45.400   &   -                           &   0.7~$[0.1,1.7]$                     &   "                                   &   \\
    6.3     &   06:00:10.646    &   -20:06:52.900   &   -                           &   0.6~$[0.2,2.3]$                     &   "                                   &   \\
    6.4     &   06:00:10.705    &   -20:06:50.380   &   -                           &   -                                   &   "                                   &   \\\hline
    7.1     &   06:00:09.258    &   -20:08:25.725   &   $6.0719\pm0.0004$           &   -                                   &   -                                   &   Multiply imaged ALMA system\\
    7.2     &   06:00:09.002    &   -20:08:27.381   &   "                           &   -                                   &   -                                   &   \citep{laporte21b,fujimoto21}.\\
    7.3     &   06:00:08.562    &   -20:08:12.260   &   "                           &   -                                   &   -                                   &   "; Analyzed in \citet{fujimoto24}.\\
    7.4     &   06:00:09.554    &   -20:08:10.749   &   "                           &   -                                   &   -                                   &   \\
    7.5     &   06:00:05.305    &   -20:07:27.646   &   -                           &   -                                   &   -                                   &   Counter-image confirmed with JWST.\\\hline
    8.1     &   06:00:08.027    &   -20:08:24.975   &   $5.4603\pm0.0001$           &   -                                   &   -                                   &   \\
    8.2     &   06:00:10.317    &   -20:08:12.355   &   $5.4599\pm0.0001$           &   -                                   &   -                                   &   \\
    8.3     &   06:00:10.146    &   -20:08:01.761   &   $5.4603\pm0.0001$           &   -                                   &   -                                   &   \\\hline
    9.1     &   06:00:10.199    &   -20:08:11.399   &   $3.521\pm0.001$             &   -                                   &   -                                   &   \\
    9.2     &   06:00:08.103    &   -20:08:23.457   &   $3.5247\pm0.0003$           &   -                                   &   -                                   &   \\
    9.3     &   06:00:10.073    &   -20:08:02.763   &   $3.5169\pm0.0009$           &   -                                   &   -                                   &   \\
    9.4     &   06:00:08.235    &   -20:08:11.677   &   $3.5228\pm0.0003$           &   -                                   &   -                                   &   \\\hline
    10.1    &   06:00:08.616    &   -20:08:13.016   &   $4.5064\pm0.0001$           &   -                                   &   -                                   &   \\
    10.2    &   06:00:09.470    &   -20:08:14.156   &   $4.5066\pm0.0001$           &   -                                   &   -                                   &   \\
    10.3    &   06:00:09.160    &   -20:08:24.086   &   $4.5071\pm0.0002$           &   -                                   &   -                                   &   \\\hline
    11.1    &   06:00:10.312    &   -20:08:09.741   &   $3.6753\pm0.0001$           &   -                                   &   -                                   &   \\
    11.2    &   06:00:10.212    &   -20:08:02.284   &   $3.6753\pm0.0001$           &   -                                   &   -                                   &   \\\hline
    12.1    &   06:00:10.224    &   -20:07:49.108   &   $5.1734\pm0.0001$           &   -                                   &   -                                   &   \\
    12.2    &   06:00:10.928    &   -20:08:06.870   &   $5.1739\pm0.0002$           &   -                                   &   -                                   &   \\\hline
    13.1    &   06:00:10.470    &   -20:08:32.123   &   $3.0292\pm0.0002$           &   -                                   &   -                                   &   \\
    13.2    &   06:00:08.533    &   -20:07:48.574   &   $3.0292\pm0.0003$           &   -                                   &   -                                   &   \\\hline
    14.1    &   06:00:09.807    &   -20:08:35.307   &   $4.4406\pm0.0003$           &   -                                   &   -                                   &   \\
    14.2    &   06:00:08.977    &   -20:07:56.741   &   $4.4404\pm0.0002$           &   -                                   &   -                                   &   \\\hline
    15.1    &   06:00:07.319    &   -20:08:05.092   &   $3.9478\pm0.0006$           &   -                                   &   -                                   &   \\
    15.2    &   06:00:07.491    &   -20:08:04.934   &   $3.947\pm0.003$             &   -                                   &   -                                   &   \\
    15.3    &   06:00:11.100    &   -20:08:34.476   &   $3.945\pm0.004$             &   -                                   &   -                                   &   \\\hline
    16.1    &   06:00:09.037    &   -20:08:05.542   &   $4.1399\pm0.0008$           &   -                                   &   -                                   &   Radial system; close to south BCG.\\
    16.2    &   06:00:08.604    &   -20:08:08.207   &   $4.1326\pm0.0002$           &   -                                   &   -                                   &   "\\\hline
    17.1    &   06:00:09.727    &   -20:07:57.920   &   $3.32\pm0.05$               &   -                                   &   -                                   &   \\
    17.2    &   06:00:10.296    &   -20:08:19.373   &   $3.3162\pm0.0001$           &   -                                   &   -                                   &   \\
    17.3    &   06:00:07.755    &   -20:08:23.942   &   $3.3162\pm0.0001$           &   -                                   &   -                                   &   \\\hline
    \end{tabular}
    \par\smallskip
    \begin{justify}
        \texttt{Note.} -- \textit{Column~1}: Image ID. -- \textit{Columns~2 and~3}: Right ascension and declination. -- \textit{Column~4}: Spectroscopic redshift. -- \textit{Column~5}: Photometric redshift and its 95\,\% ($2\sigma$) range as computed with \texttt{BPZ} in the RELICS catalog (see section~\ref{sec:RELICS}). -- \textit{Column~6}: Median redshift from our SL-model (see section~\ref{sec:SL_result}). -- \textit{Column~7}: Additional remarks.
    \end{justify}
    \label{tab:multiple_images}
\end{table*}

\begin{table*}
    \contcaption{Multiple images and candidates identified in the Anglerfish cluster.}
    \begin{tabular}{lcccccr}
    \hline
    ID      &   RA              &   Dec.            &   $z_{\mathrm{spec}}$         &   $z_{\mathrm{phot}}$~[95\,\%-range]  &   $z_{\mathrm{model}}$~[95\,\%-range] &   Remarks\\\hline
    18.1    &   06:00:08.078    &   -20:08:20.160   &   $3.6672\pm0.0003$           &   -                                   &   -                                   &   \\
    18.2    &   06:00:10.312    &   -20:08:09.741   &   $3.674\pm0.001$             &   -                                   &   -                                   &   \\\hline    
    19.1    &   06:00:07.901    &   -20:07:49.735   &   $3.9500\pm0.0004$           &   -                                   &   -                                   &   \\
    19.2    &   06:00:07.877    &   -20:08:04.171   &   $3.9370\pm0.0002$           &   -                                   &   -                                   &   \\\hline
    20.1    &   06:00:07.966    &   -20:07:52.637   &   $5.3550\pm0.0001$           &   -                                   &   -                                   &   \\
    20.2    &   06:00:10.913    &   -20:08:39.110   &   $5.3549\pm0.0006$           &   -                                   &   -                                   &   \\\hline
    21.1    &   06:00:06.799    &   -20:08:15.872   &   -                           &   -                                   &   -                                   &   \\
    21.2    &   06:00:08.583    &   -20:07:46.257   &   $5.4086\pm0.0002$           &   -                                   &   -                                   &   \\
    21.3    &   06:00:10.745    &   -20:08:33.675   &   $5.4087\pm0.0001$           &   -                                   &   -                                   &   \\\hline
    22.1    &   06:00:08.994    &   -20:07:45.452   &   $3.526\pm0.001$             &   -                                   &   -                                   &   \\
    22.2    &   06:00:10.718    &   -20:08:24.385   &   $3.529\pm0.004$             &   -                                   &   -                                   &   \\
    22.3    &   06:00:07.392    &   -20:08:18.134   &   $3.529\pm0.001$             &   -                                   &   -                                   &   \\\hline
    23.1    &   06:00:07.477    &   -20:07:39.701   &   $5.4087\pm0.0002$           &   -                                   &   -                                   &   \\
    23.2    &   06:00:07.709    &   -20:07:40.085   &   $5.4090\pm0.0003$           &   -                                   &   -                                   &   \\\hline
    c24.1   &   06:00:20.084    &   -20:06:40.052   &   -                           &   -                                   &   2.62~$[2.55,2.70]$                  &   System identified in the GMOS imaging.\\
    c24.2   &   06:00:20.076    &   -20:06:41.400   &   -                           &   -                                   &   "                                   &   "\\
    c24.3   &   06:00:19.472    &   -20:07:40.794   &   -                           &   -                                   &   "                                   &   "\\\hline
    c25.1   &   06:00:20.667    &   -20:06:58.993   &   -                           &   -                                   &   -                                   &   Close to BCG1; possible giant arc.\\
    c25.2   &   06:00:20.673    &   -20:07:01.917   &   -                           &   -                                   &   -                                   &   "; System identified in the GMOS imaging.\\
    c25.3   &   06:00:17.208    &   -20:06:10.908   &   -                           &   -                                   &   -                                   &   \\\hline
    26.1    &   06:00:05.591    &   -20:08:54.752   &   -                           &   1.8~$[1.2,2.0]$                     &   2.14~$[1.99,2.45]$                  &   Galaxy-galaxy system.\\
    26.2    &   06:00:05.303    &   -20:08:51.220   &   -                           &   1.1~$[0.7,1.6]$                     &   "                                   &   \\
    26.3    &   06:00:05.398    &   -20:08:53.966   &   -                           &   -                                   &   "                                   &   \\
    26.4    &   06:00:05.453    &   -20:08:52.327   &   -                           &   1.0~$[0.9,1.1]$                     &   "                                   &   \\\hline
    27.1    &   06:00:07.351    &   -20:08:07.433   &   -                           &   -                                   &   7.01~$[6.70,7.25]$                  &   System identified in JWST imaging.\\
    27.2    &   06:00:07.591    &   -20:08:06.868   &   -                           &   -                                   &   "                                   &   "\\\hline
    28.1    &   06:00:08.897    &   -20:08:06.337   &   -                           &   -                                   &   2.76~$[2.72,2.92]$                  &   Radial system; close to south BCG.\\
    28.2    &   06:00:08.817    &   -20:08:08.057   &   -                           &   1.8~$[1.4,2.5]$                     &   "                                   &   "\\\hline
    c29.1   &   06:00:08.917    &   -20:08:08.118   &   -                           &   -                                   &   -                                   &   Radial system; close to south BCG.\\
    c29.2   &   06:00:08.531    &   -20:08:09.482   &   -                           &   -                                   &   -                                   &   "\\\hline
    30.1    &   06:00:05.602    &   -20:08:35.526   &   -                           &   -                                   &   6.69~$[6.26,7.10]$                  &   System identified in JWST imaging.\\
    30.2    &   06:00:05.744    &   -20:08:38.406   &   -                           &   -                                   &   "                                   &   "\\\hline
    c31.1    &   06:00:14.186    &   -20:07:29.545   &   -                          &   0.5~$[0.2,2.5]$                     &   1.06~$[1.05,1.10]$                  &   \\
    c31.2    &   06:00:14.159    &   -20:07:27.857   &   -                          &   2.5~$[1.4,2.8]$                     &   "                                   &   \\\hline
    c32.1    &   06:00:13.784    &   -20:07:29.722   &   -                          &   -                                   &   3.78~$[3.76,3.80]$                  &   \\
    c32.2    &   06:00:13.737    &   -20:07:27.920   &   -                          &   -                                   &   "                                   &   \\\hline
    c33.1    &   06:00:16.581    &   -20:06:38.658   &   -                          &   -                                   &   1.05~$[1.03,1.06]$                  &   System identified in the GMOS imaging.\\
    c33.2    &   06:00:16.958    &   -20:06:46.349   &   -                          &   -                                   &   "                                   &   "\\
    c33.3    &   06:00:17.301    &   -20:07:21.123   &   -                          &   -                                   &   "                                   &   "\\\hline
    c34.1   &   06:00:20.142    &   -20:07:10.303   &   -                           &   -                                   &   -                                   &   Faint and diffuse, only detected in the \textit{i}-band.\\
    c34.2   &   06:00:17.046    &   -20:06:13.842   &   -                           &   -                                   &   -                                   &   "; System identified in the GMOS imaging.\\
    c34.3   &   06:00:18.269    &   -20:07:38.249   &   -                           &   -                                   &   -                                   &   "\\\hline
    c35.1   &   06:00:13.452    &   -20:07:06.246   &   -                           &   1.2~$[1.1,1.3]$                     &   -                                   &   At the eastern edge of the HST imaging.\\
    c35.2   &   06:00:20.841    &   -20:06:50.397   &   -                           &   -                                   &   -                                   &   Detected in the GMOS imaging.\\\hline
    \end{tabular}
    \par\smallskip
    \begin{justify}
        \texttt{Note.} -- \textit{Column~1}: Image ID. -- \textit{Columns~2 and~3}: Right ascension and declination. -- \textit{Column~4}: Spectroscopic redshift. -- \textit{Column~5}: Photometric redshift and its 95\,\% ($2\sigma$) range as computed with \texttt{BPZ} in the RELICS catalog (see section~\ref{sec:RELICS}). -- \textit{Column~6}: Median redshift from our SL-model (see section~\ref{sec:SL_result}). -- \textit{Column~7}: Additional remarks.
    \end{justify}
\end{table*}

The multiple images and candidates identified in MACS0600 are all listed in Tab.~\ref{tab:multiple_images}, and shown in Figs.~\ref{fig:multiple-images}. The spectroscopic redshfift measurement of system 7 comes from the ALMA data and was first published in \citet{fujimoto21}. Other spectroscopic redshifts were measured with MUSE as described in section~\ref{sec:muse} \citep[for the method, see also][]{richard21}. Most of the MUSE multiple image spectroscopic redshifts are based on the Lyman-$\alpha$ line, apart for systems 1, 2, 3, 10, 18 and 19 which are based on absorption lines and forbidden emission lines. All MUSE spectroscopic redshifts shown in Tab.~\ref{tab:multiple_images} have confidence levels of 2 or 3. A machine-readable version of Tab.~\ref{tab:multiple_images} is available in the online supplementary material of this paper and contains further information: cross-matching to the public RELICS catalog and the MUSE catalog that will be made publicly available on the \citet[][]{richard21} Atlas of MUSE observations towards massive lensing clusters website, and flagging if used in our model or not.


\bsp	
\label{lastpage}
\end{document}